\documentclass[twocolumn,showpacs,superscriptaddress,floatfix,longbibliography,prx]{revtex4-2}
\usepackage{graphicx,amsfonts,amssymb,amsmath,hyperref,hypcap,enumerate}
\usepackage{amsthm}
\usepackage{multirow}
\usepackage{graphicx}
\usepackage{dcolumn}   
\usepackage{bm}        
\usepackage{amssymb}   
\usepackage{amsmath}
\usepackage{epstopdf}
\usepackage{color}
\usepackage{subfigure}
\usepackage{diagbox}

\usepackage{multirow}
\usepackage{ bbold }

\DeclareMathAlphabet{\mathitb}{OT1}{cmr}{bx}{sl}

\begin{document}
	\title{Transfer matrix study of the Anderson transition in non-Hermitian systems}
	\author{Xunlong Luo}
	\email{luoxunlong@pku.edu.cn}
	\affiliation{Science and Technology on Surface Physics and Chemistry Laboratory, Mianyang 621907, China}
	
	\author{Tomi Ohtsuki}
	\email{ohtsuki@sophia.ac.jp}
	\affiliation{Physics Division, Sophia University, Chiyoda-ku, Tokyo 102-8554, Japan}
	
	\author{Ryuichi Shindou}
	\email{rshindou@pku.edu.cn}
	\affiliation{International Center for Quantum Materials, Peking University, Beijing 100871, China}
	\affiliation{Collaborative Innovation Center of Quantum Matter, Beijing 100871, China}
	
	\date{\today}
	\begin{abstract}
		The Anderson transition driven by non-Hermitian (NH) disorder has been 
		extensively studied in recent years. In this paper, we present in-depth transfer 
		matrix analyses of the Anderson transition in three 
		NH systems, NH Anderson, U(1), and Peierls models in three-dimensional systems. 
		The first model belongs to NH class AI$^{\dagger}$, whereas the second and the third ones
		to NH class A.
		We first argue a general validity of the transfer matrix analysis in NH systems, 
		and clarify the symmetry properties of the Lyapunov exponents, scattering ($S$) matrix and 
		two-terminal conductance in these NH models. The unitarity of the $S$ matrix is violated in 
		NH systems, where the two-terminal conductance can take arbitrarily large values. Nonetheless, we 
		show that the transposition symmetry of a Hamiltonian leads to the symmetric nature of the $S$ matrix as well as 
		the reciprocal symmetries of the Lyapunov exponents and conductance in certain ways in these NH 
		models. Using the transfer matrix 
		method, we construct a phase diagram of the NH Anderson model for various complex single-particle energy $E$. 
		At $E=0$, the phase diagram as well as critical properties become completely symmetric with respect to 
		an exchange of real and imaginary parts of on-site NH random potentials. We show 
		that the symmetric nature at $E=0$ is a general feature for any NH bipartite-lattice 
		models with the on-site NH random potentials.
		Finite size scaling data are fitted by polynomial functions, from which
		we determine the critical exponent $\nu$ at different single-particle energies and system parameters 
		of the NH models. We conclude that the critical exponents of the NH class AI$^{\dagger}$ and the NH class A are  
		$\nu=1.19 \pm 0.01$ and $\nu=1.00 \pm 0.04$, respectively. In the NH models, a distribution of the 
		two-terminal conductance is not Gaussian. Instead, it contains small fractions of huge conductance values, 
		which come from rare-event states with huge transmissions amplified by on-site NH disorders. Nonetheless, 
		a geometric mean of the conductance enables the finite-size scaling analysis. We show that the critical 
		exponents obtained from the conductance analysis are consistent with those from the localization length 
		in these three NH models.   
	\end{abstract}
	\maketitle
	\section{introduction}
	During the last several years, physics community witnessed tremendous revivals of research interests on non-Hermitian (NH) quantum physics. Nontrivial topological band theories have been introduced 
	in NH systems, including the breakdown of the bulk-boundary 
	correspondence and the emergence of new topological invariants\cite{Esaki11,Lee16,Leykam17,Kunst18,Martinez18,Yao18,Yao18_2,Gong18,Lieu18,Yin18,Yokomizo19,Deng19,Song19,Liu19},
	enriched topological classifications according to symmetries \cite{Zhou19,Kawabata19,Kawabata19NC},
	and NH skin effects \cite{Song19_2,Okuma19,Lee19,Jiang19,Ezawa19,Okuma20}.
	NH quantum phenomena occur not only in condensed matter systems, but also in photonic systems \cite{Guo09,Ruter10,Feng11,Regensburger12,Zeuner15,Malzard15,Zhen15,Weimann17,Xiao17,Bahari17,Zhou18,Bandres18,Harari18} 
	and ultracold atoms \cite{Xu17,Nakagawa18,Li19}. Many of NH quantum systems have 
	quenched disorders by their own generic nature. Canonical examples of this is 
	random lasers in a region of random dissipation and amplification~\cite{Cao99,Wiersma08,Wiersma13}, 
	nonequilibrium open systems with gain and/or loss\cite{Konotop16,Feng17,El-Ganainy18,Ozdemir19,Miri19}, 
	and correlated quantum many-particle systems of quasiparticles with finite life-time \cite{Shen18,Papaj19}.
	Nonetheless, the Anderson transition~\cite{Anderson58} in the NH quantum systems has  
	been poorly studied in higher spatial  
	dimension~\cite{Xu16,Tzortzakakis20,Wang20,Huang20,Huang20SR,Luo21}, 
	in spite of the quarter century history since the Hatano-Nelson's pioneering work of the one-dimensional (1D) 
	Anderson model~\cite{Hatano96,Kawabata20}. Hatano and Nelson introduced the 
	non-Hermiticity through asymmetric hopping terms in the 1D Anderson model. 
	Recent works focused on the effect of on-site complex-valued random 
	potentials in Anderson and U(1) models in the two and three 
	dimensions~\cite{Tzortzakakis20,Wang20,Huang20,Huang20SR,Luo21}. 
	A positive imaginary part of the on-site potential gives a gain of a wavefunction amplitude in 
	time, while a negative value of the imaginary part of the potential gives a loss of the wavefunction 
	amplitude. 
	Such complex-valued random potentials can be experimentally implemented by random dissipation and 
	amplification of light waves in random lasers~\cite{Cao99,Wiersma08,Wiersma13}.
	
	There are three major shortcomings in the previous studies of the Anderson transition in 
	the NH quantum systems. First of all, all the studies so far~\cite{Huang20,Luo21} 
	are based on eigenenergy level statistics~\cite{Wigner51,Dyson62,Dyson62TFW}, 
	which suffers from an ambiguity of how to set a window for the single-particle energy. 
	The level statistics analysis sets a finite width of the energy window over which a dimensionless 
	quantity associated with the level statistics is averaged. Thus, when using a finite-size scaling 
	analysis in favor for the universal critical exponent, the analysis with a nonlinear polynomial fitting  
	assumes that {\it any} fitting parameters in a universal function for the dimensionless quantity 
	have {\it no} variations within the energy window. 
	Nonetheless, the parameters in the universal function include not only 
	the universal critical exponent but also non-universal critical quantities such as critical disorder 
	strength. These non-universal quantities are continuous functions of the energy. Thus,  
	the energy window must be sufficiently narrow, such that all the single-particle states 
	within the window could share the same values of these non-universal quantities. To 
	take sufficient amount of the level statistics with such a narrow window, one needs to diagonalize   
	single-particle Hamiltonian with many different disorder realizations. This often meets an upper 
	limit from computational resources. 
	Large statistical errors due to limited number of samples together with the systematic errors
	due to finite energy window make it hard for the precise investigation of the Anderson transition 
	in NH quantum systems.  
	
	Eigenvalues of the NH systems are distributed in the complex 
	Euler plane, except for a system with a special symmetry, such as 
	${\cal PT}$ symmetry~\cite{Bender98}. Thereby, the complex plane can possibly 
	host a novel phase diagram structure, which has no counterpart in the 
	Hermitian physics. To uncover such unique NH quantum physics, one needs to clarify how critical 
	nature of single-particle states varies as a function of their complex-valued eigenenergy.
	It is also unclear how the known universality classes of the Anderson transition in the 
	Hermitian case cross over to new universality classes in the non-Hermitian case. To clarify this, 
	one needs to study precisely how the critical properties change as a function of the system parameters. 
	
	The two-terminal conductance is one of the best physical quantities that characterize the Anderson 
	transition in the experimental NH systems mentioned above. Nonetheless, it is unknown how the conductance 
	as well as the localization length behave in localized, delocalized phases and at the critical point 
	in the NH disordered systems. To know them, one needs a transfer matrix analyses of the conductance 
	in the NH systems. In the NH systems, however, a symplectic structure of the transfer matrix is 
	absent and so is the unitarity of a scattering matrix. Besides, the presence of the reciprocal 
	symmetries of the conductance and Lyapunov exponents is not clear in the NH quantum systems. In fact, 
	the reciprocal symmetry is completely absent in some NH systems such as in the 1D Hatano-Nelson model, 
	while it exists in other NH systems in some ways.

	In order to solve these shortcomings in theory of NH quantum systems, we carry out comprehensive 
	transfer matrix analyses of localization length and conductance in the three-dimensional (3D) NH Anderson, 
	U(1) and Peierls models. Here we list the major findings in this paper as follows,
	
	\begin{figure*}[tb]
		\centering
		\subfigure[Phase diagram at $E=0$]{
			\begin{minipage}[t]{0.33\linewidth}
				\centering
				\includegraphics[width=1\linewidth]{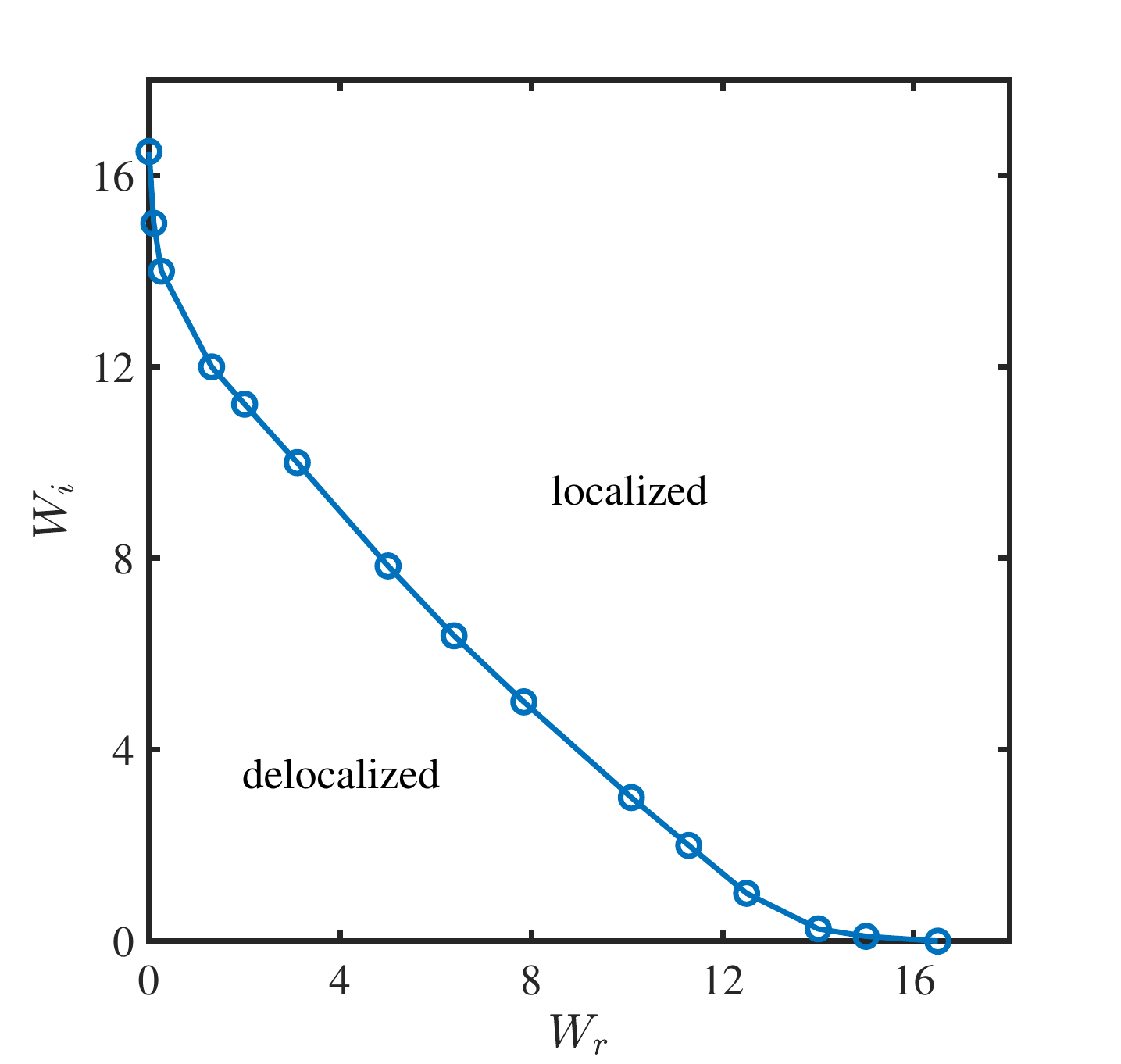}
			\end{minipage}%
		}%
		\subfigure[ RG flow at $E=0$ ]{
			\begin{minipage}[t]{0.33\linewidth}
				\centering
				\includegraphics[width=0.95\linewidth]{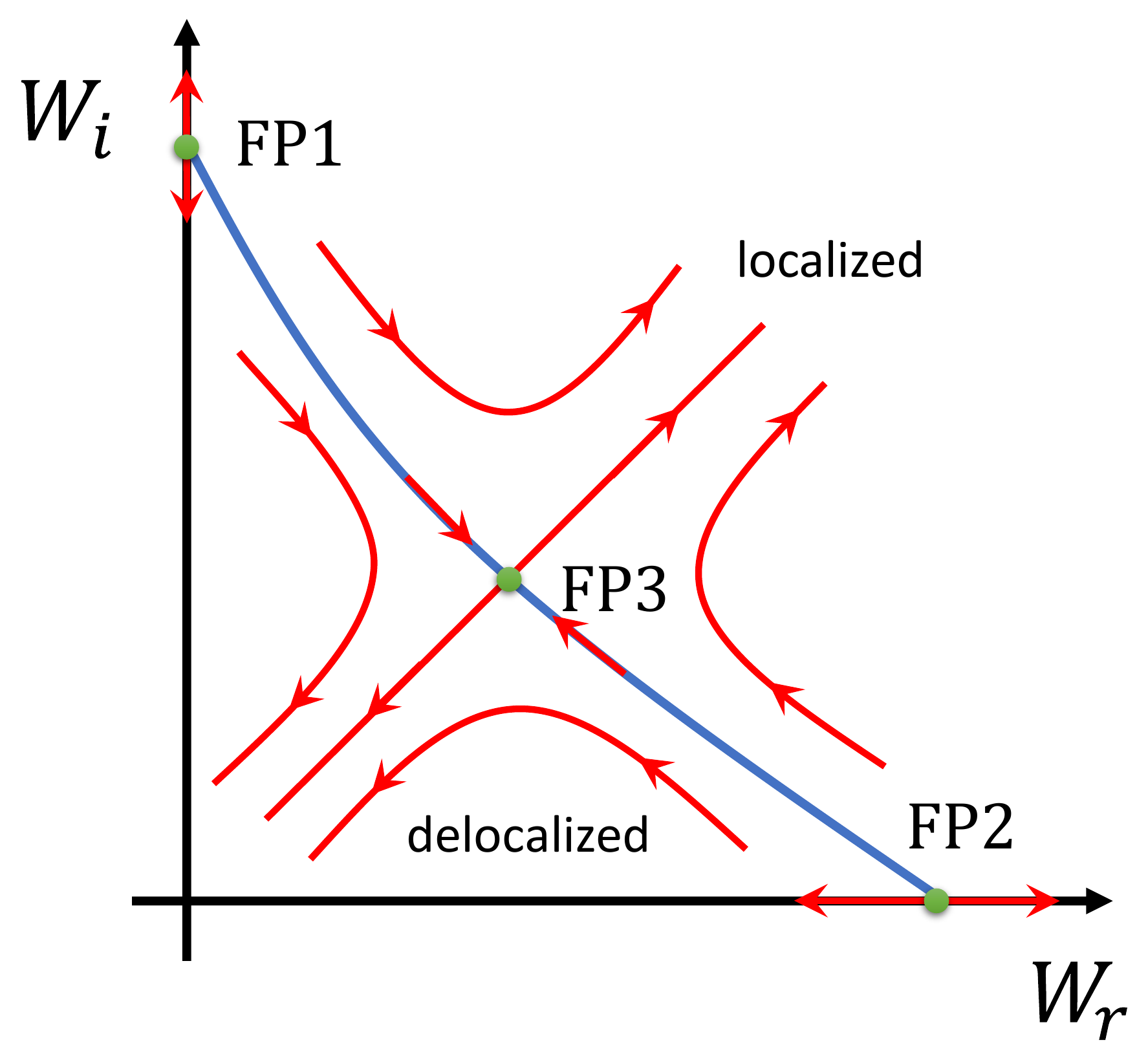}
			\end{minipage}%
		}%
		\subfigure[ RG flow at $E\ne0$ ]{
			\begin{minipage}[t]{0.33\linewidth}
				\centering
				\includegraphics[width=0.95\linewidth]{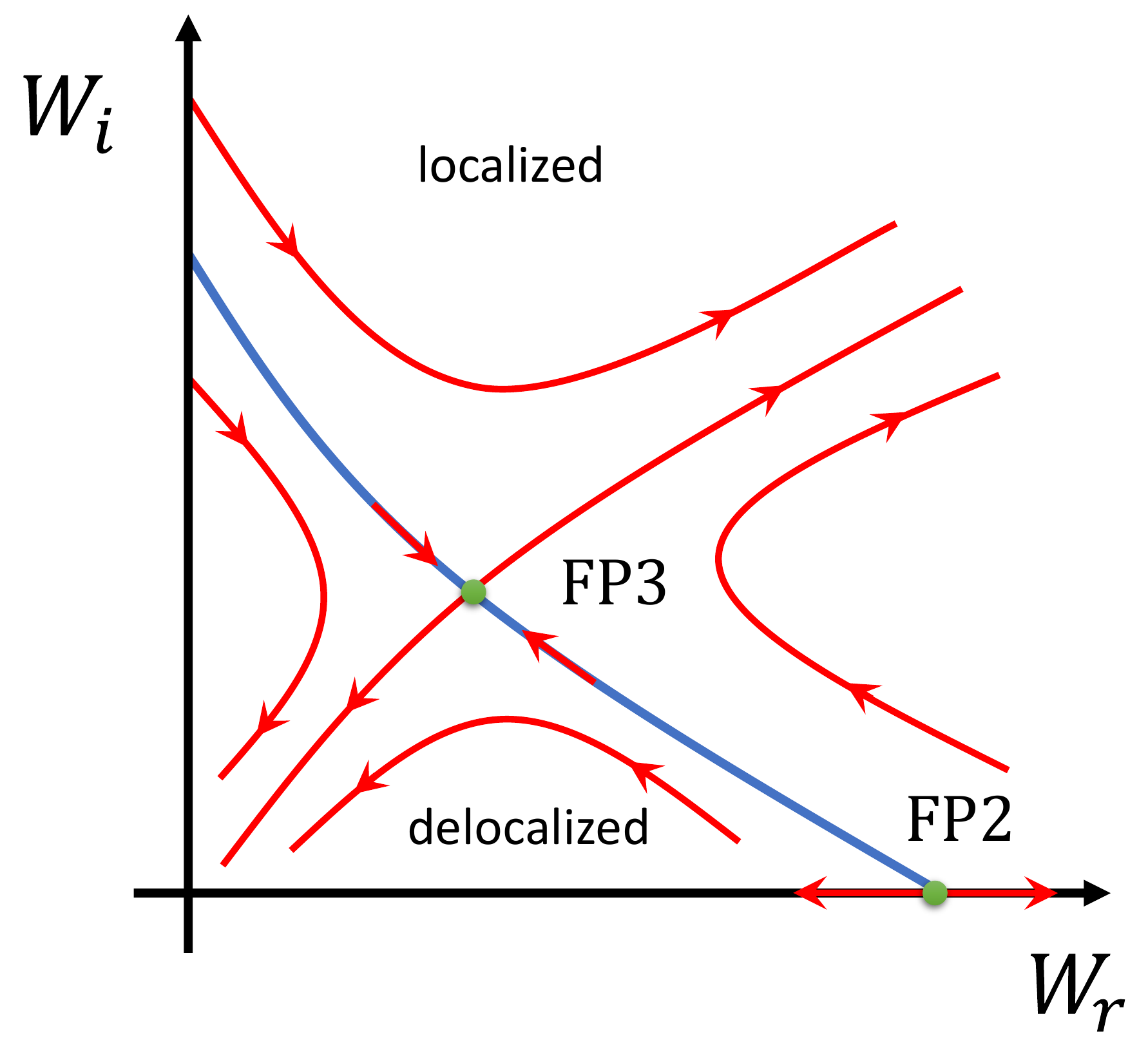}
			\end{minipage}%
		}%
		\caption{ (a) $E=0$ phase diagram of the NH Anderson model   
			determined by the localization length, (b) $E=0$ and (c) $E\ne0$ renormalization group (RG) 
			flow diagram for the NH Anderson model, where $W_r$ and $W_i$ are the 
			disorder strength for the real and imaginary parts of the complex-valued 
			on-site random potentials, respectively. In (b) and (c), the red curves with arrows stand for the 
			RG flow, the green dots for fixed points (FP) of the RG flow, and the blue line is a phase boundary 
			between delocalized and localized phases.  Note that the phase boundary and the RG flow is completely 
			symmetric about a $W_r=W_i$ line in (a) and (b).
			We also note that $E\ne 0$ in (c) means that $E$ is real.} 
		\label{phasediagram}
	\end{figure*}
	
	\begin{itemize}
		\item {The critical exponents of the NH Anderson, U(1) and Peierls models are 
			accurately determined. The results unambiguously conclude that 
			the critical exponent of the NH class-AI$^{\dagger}$ (Anderson) is $\nu=1.19 \pm 0.01$ 
			and the critical exponent 
			of the NH class A (U(1) and Peierls) is $\nu=1.00 \pm 0.04$.
			These high precision estimates enable us to distinguish 
			the NH class A and NH class AI$^\dagger$, which was impossible 
			for the previous level statistics study~\cite{Luo21}. 
			This supports a strong relation between the universality 
			classes of the Anderson transition and symmetry classification in the NH disordered systems. 
			More importantly, they are distinctly different from the critical exponents in 
			the corresponding universality classes in the Hermitian limit; $\nu= 1.572 \pm 0.003 $ in the 
			orthogonal class~\cite{Slevin18} (Hermitian class AI)
			and $\nu=1.443 \pm 0.003 $ in unitary class (Hermitian class A)
			~\cite{Kawarabayashi98,Slevin16}.}   
		\item{Critical properties at different single-particle energy $E$ and system parameters are clarified in the NH Anderson 
			model. This gives phase diagrams in a two-dimensional plane subtended by the disorder strength of the real part 
			of the on-site random potential ($W_r$) and that of the imaginary part ($W_i$). Especially at the zero single-particle 
			energy ($E=0$), the phase diagram as well as the critical properties across the phase boundary becomes completely 
			symmetric with respect to an exchange of $W_r$ and $W_i$; Fig.~\ref{phasediagram}(a). The symmetric structure at $E=0$ 
			is a generic feature in any NH bipartite-lattice models with the on-site NH random potentials. Based on the results, we propose a renormalization group  (RG) flow diagram in the two-dimensional plane at 
			$E=0$; Fig.~\ref{phasediagram}(b). At $E\ne 0$, the symmetry of the phase diagram and the critical 
			properties is absent; see the RG flow diagram in Fig.~\ref{phasediagram}(c).}
		\item{The validity of the transfer matrix analyses in these 3D NH models as well as  
			the 1D Hatano-Nelson model is clarified and demonstrated numerically. The presence or absence 
			of the reciprocal symmetries in Lyapunov exponents and conductance are clarified in the three 
			NH models. Based on a physically reasonable definition of the two-terminal conductance, we show numerically 
			that a conductance distribution is not Gaussian, but it contains small numbers of extremely large conductance 
			values due to rare events of strong amplifications by the NH disorders. We also demonstrate that the geometric mean of 
			the conductance enables a feasible finite-size scaling (FSS) analysis and the results from the FSS analysis give the 
			critical exponents consistent with those values by the localization length.}
	\end{itemize}
	The organization of this paper is as follows. In the following section, we introduce the three NH models and argue 
	the validity of the transfer matrix analyses. We show the presence or absence of the reciprocal 
	symmetries in the Lyapunov exponents and two-terminal conductance. In Sec. \ref{sec:NHAM}, we clarify 
	the critical properties of the Anderson transition at different single-particle energies and system parameters in 
	the NH Anderson models together with an accurate estimate of the critical exponent of the NH 
	class AI$^{\dagger}$. In Sec. \ref{sec:NHU1}, we clarify the critical properties of the Anderson transition 
	in two NH class A models, i.e., U(1) and Peierls models, together with an estimate of the critical exponent of the 
	NH class A. 
	In Sec. \ref{sec:RLSA}, we compare the transfer matrix analyses with the level statistics 
	analyses\cite{Huang20,Luo21} and describe a possible reason for a discrepancy of the critical 
	exponent of the NH class AI$^\dagger$ between the two analyses.
	The final section is devoted to summary and concluding remarks.

	\section{numerical method}
	\label{sec:method}
	We study the following tight-binding models defined on a 3D
	cubic lattice,
	\begin{align}
		{\cal H} = \sum_{\bm i} \varepsilon_{\bm i} c^{\dagger}_{\bm i} c_{\bm i} + \sum_{\langle {\bm i},{\bm j}\rangle} 
		e^{2\pi {\rm i} \theta_{{\bm i},{\bm j}}} c^{\dagger}_{\bm i} c_{\bm j} \equiv \sum_{{\bm i},{\bm j}} c^{\dagger}_{\bm i} (\mathbb{H})_{{\bm i},{\bm j}} c_{\bm j},  
		\label{TBmodel}
	\end{align}
	where $c^{\dagger}_{\bm i}$ ($c_{\bm i}$) is the creation (annihilation) operator. $\bm i$ and $\bm j$  
	specify the cubic lattice site; ${\bm i}=(i_x,i_y,i_z)$ and $i_{\mu}=1,2,\cdots,L_{\mu}$ 
	($\mu=x,y,z$). The models have   
	Hermitian hoppings between nearest neighbor sites in the cubic lattice; $\langle {\bm i},{\bm j}\rangle$ means 
	that $\bm i$ and $\bm j$ are the nearest neighbor sites with $\theta_{{\bm i},{\bm j}} = -\theta_{{\bm j},{\bm i}}$. 
	The model is the Anderson model for $\theta_{{\bm i},{\bm j}}$ = 0 and the U(1) model with a random 
	number $\theta_{{\bm i},{\bm j}}$ in a range of $[0, 1)$ \cite{Ohtsuki94,Kawarabayashi98}.  When $\theta_{{\bm i},{\bm j}} = \Phi \!\ i_x$ 
	for ${\bm j}={\bm i}+{\bm e}_z$, the model is the Peierls model \cite{Peierls33,Luttinger51}, where $2\pi \Phi$ is a magnetic gauge flux that 
	penetrates through every square plaquette in the $z$-$x$ plane of the cubic lattice. 
	In this paper, we study the Anderson transition driven by random complex on-site potentials, 
	$\varepsilon_{\bm i}  = w^r_{\bm i} + {\rm i}\!\ w^i_{\bm i}$ with the imaginary unit ${\rm i}$, where 
	$w^r_{\bm i}$ and $w^i_{\bm i}$ are independent random numbers with the uniform distribution 
	in a range of $[-W_r/2,W_r/2]$ and $[-W_i/2,W_i/2]$, respectively. Hence 
	${\cal H}^{\dagger}\ne {\cal H}$, where the non-zero $w^r_{\bm i}$ and $w^i_{\bm i}$ 
	bring about the non-Hermiticity in the system. 
	According to the symmetry classification, the Anderson model ($\theta_{{\bm i},{\bm j}}=0$) 
	belongs to 3D NH class AI$^{\dagger}$ and 
	the U(1) and Peierls models ($\theta_{{\bm i},{\bm j}} \ne 0$) belong to 3D NH class A. The time reversal 
	symmetry (TRS) is broken (${\cal H}^{*} \ne {\cal H}$) in  both classes, whereas the 
	transposition symmetry (${\cal H}^\mathrm{T} = {\cal H}$), namely TRS$^{\dagger}$, holds true
	in the class AI$^{\dagger}$\cite{Kawabata19,Hamazaki20}. We note that not only a parameter region of 
	$W_i=0$ but also a parameter region of $W_r=0$ and $E=0$ belong to the Hermitian universality 
	class in these three NH models (Sec.~\ref{sec:NHAM}).

Transfer matrix method are widely used both in the Hermitian~\cite{MacKinnon81,Pichard81,MacKinnon83,Slevin14} and non-Hermitian~\cite{Artoni05,Mostafazadeh09,Longhi10,Ge12,Reddy13,Ramezani14,Basiri15,Loran16,Mullers18,Konotop19,Simon19} systems.
In order to study the Anderson transition of 
eigenstates of ${\cal H}$ with a complex-valued eigenenergy $E$, the quasi-1D 
localization length (Lyapunov exponent) and two-terminal conductance in 
cubic systems are evaluated by the transfer matrix method at the energy $E$ for 
different system sizes. The quantities are further analyzed by the finite 
size scaling, which determines the critical properties of 
the Anderson transition in the NH systems. Although the transfer matrix method has been 
widely used in the non-Hermitian optical systems during last several decades, 
the transfer matrix analysis on the Anderson transitions in non-Hermitian 
systems has not been carried out prior to this work. Thus, we first clarify the 
nature of the Lyapunov exponent and the two-terminal conductance in the 
NH systems studied in this paper as well as pseudo-Hermitian systems.  
	
	\subsection{transfer matrix method}
	In the transfer matrix method, the 3D $L_x \times L_y \times L_z (=N)$ cubic 
	lattice is regarded as a multiple-layer structure of its two-dimensional (2D) slices, 
	where the 2D slice is in the $x$-$y$ plane of the cubic lattice and the slices are stacked 
	along $z$. An eigenvalue problem of the $N \times N$ disordered 
	NH Hamiltonian reduces to a set of linear equations relating the 
	wavefunction on the adjacent slices. The equation takes 
	a form of a matrix multiplication of a transfer matrix that connects the wavefunctions 
	on the adjacent slices\cite{MacKinnon81,Pichard81,MacKinnon83,Kramer93,MacKinnon94,Slevin14},
	\begin{equation}
		\begin{pmatrix} \psi_{i_z+1}\\ V_{i_z+1,i_z}\psi_{i_z}\end{pmatrix}=
		M_{i_z}\begin{pmatrix} \psi_{i_z}\\ V_{i_z,i_z-1}\psi_{i_z-1}\end{pmatrix}, \label{eq2}
	\end{equation}
	where $\psi_{i_z}$ is the wavefunction on the slice at position $i_z$. 
	$M_{i_z}$ is the transfer matrix defined by,
	\begin{equation}
		M_{i_z}=\begin{pmatrix} V_{i_z,i_z+1}^{-1}(E-H_{i_z}) &-V_{i_z,i_z+1}^{-1}\\ V_{i_z+1,i_z}& 0\end{pmatrix}, \label{eqn2a}
	\end{equation} 
	$V_{i_z+1,i_z}$ is a hopping term matrix between the slice at $i_z+1$ and that 
	at $i_z$. In Eq.~(\ref{TBmodel}), $V_{i_z+1,i_z}$ is a $L_x L_y$ by $L_x L_y$ diagonal matrix 
	whose diagonal elements have the unit modulus. Note that for NH U(1) model,  
	we use local gauge transformation to eliminate the phase of the hopping along $z$-direction.
	$H_{i_z}$ consists of a hopping term and diagonal term within the slice at $i_z$. In Eq.~(\ref{TBmodel}), 
	the diagonal elements of $H_{i_z}$ are the complex-valued on-site potentials, and off-diagonal 
	elements are the nearest neighbor hoppings within the slice. The localization length 
	and two-terminal conductance are calculated along $z$ with either periodic boundary 
	conditions (PBC) or open boundary conditions (OBC) in the transverse direction, $x$ and $y$.

	\subsubsection{localization length}
	
	\begin{figure*}[tb]
		\centering
		\subfigure[Anderson model ]{
			\begin{minipage}[t]{0.2\linewidth}
				\centering
				\includegraphics[width=1\linewidth]{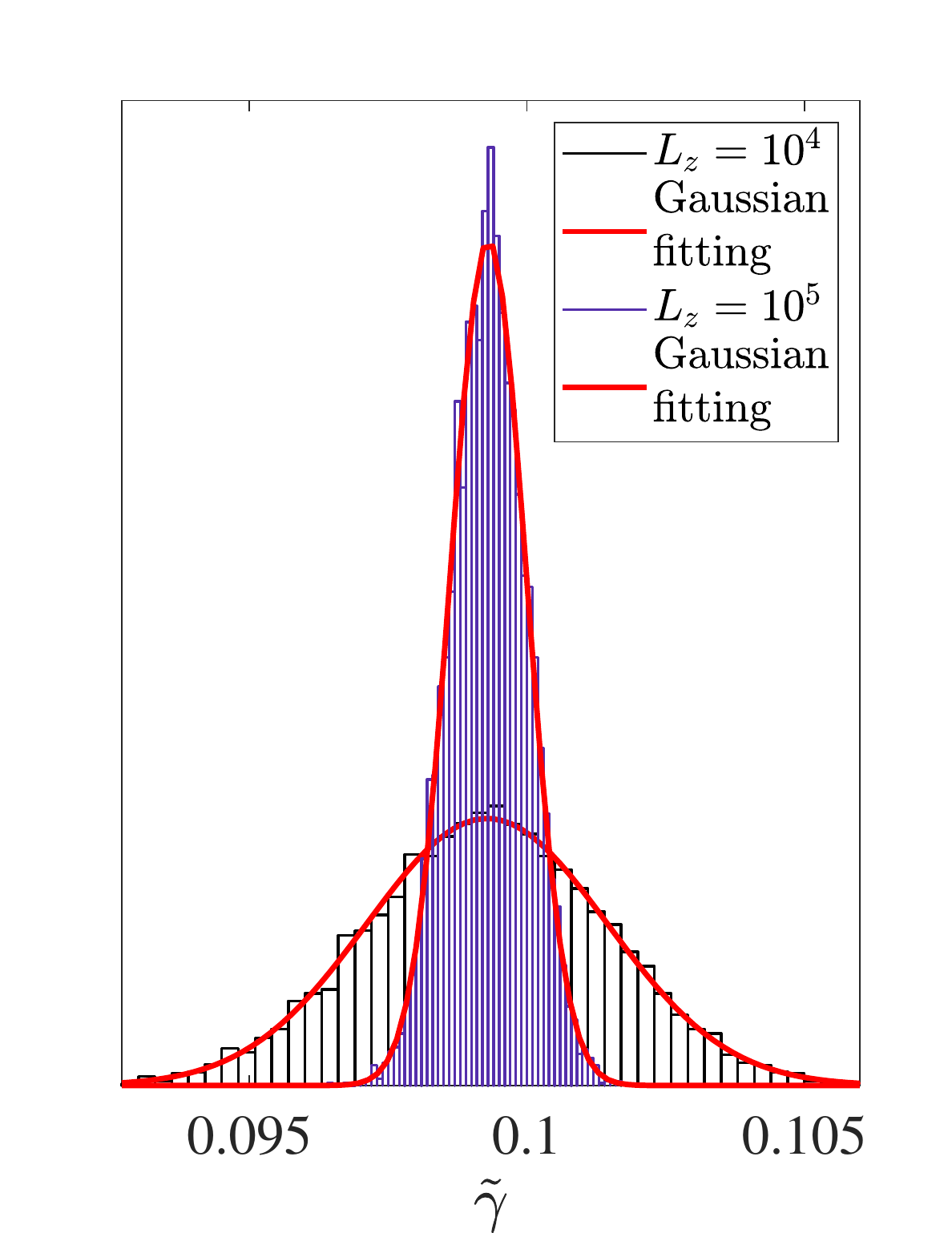}
			\end{minipage}%
		}%
		\subfigure[ U(1) model]{
			\begin{minipage}[t]{0.2\linewidth}
				\centering
				\includegraphics[width=1\linewidth]{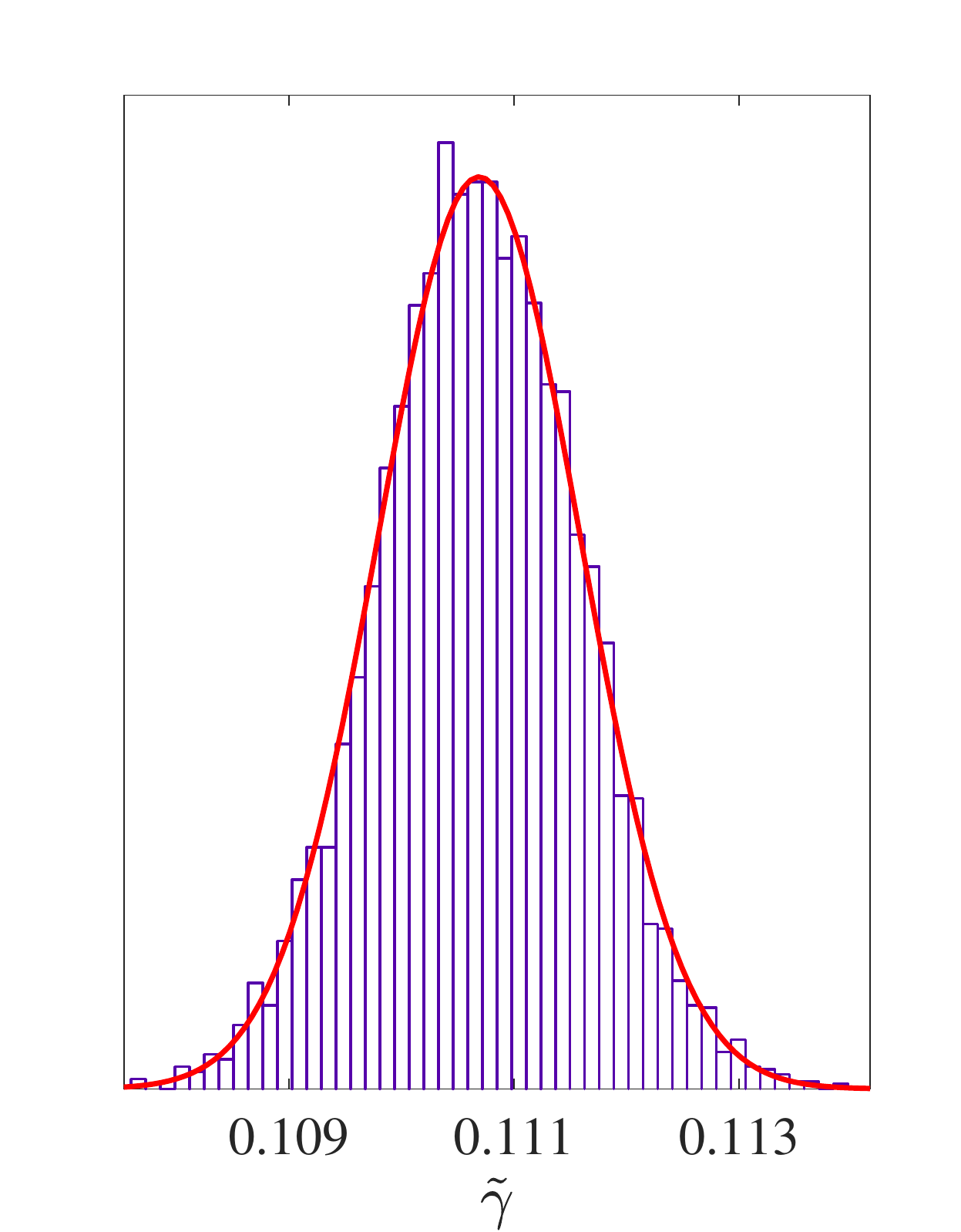}
			\end{minipage}%
		}%
		\subfigure[Hatano-Nelson model]{
			\begin{minipage}[t]{0.2\linewidth}
				\centering
				\includegraphics[width=1\linewidth]{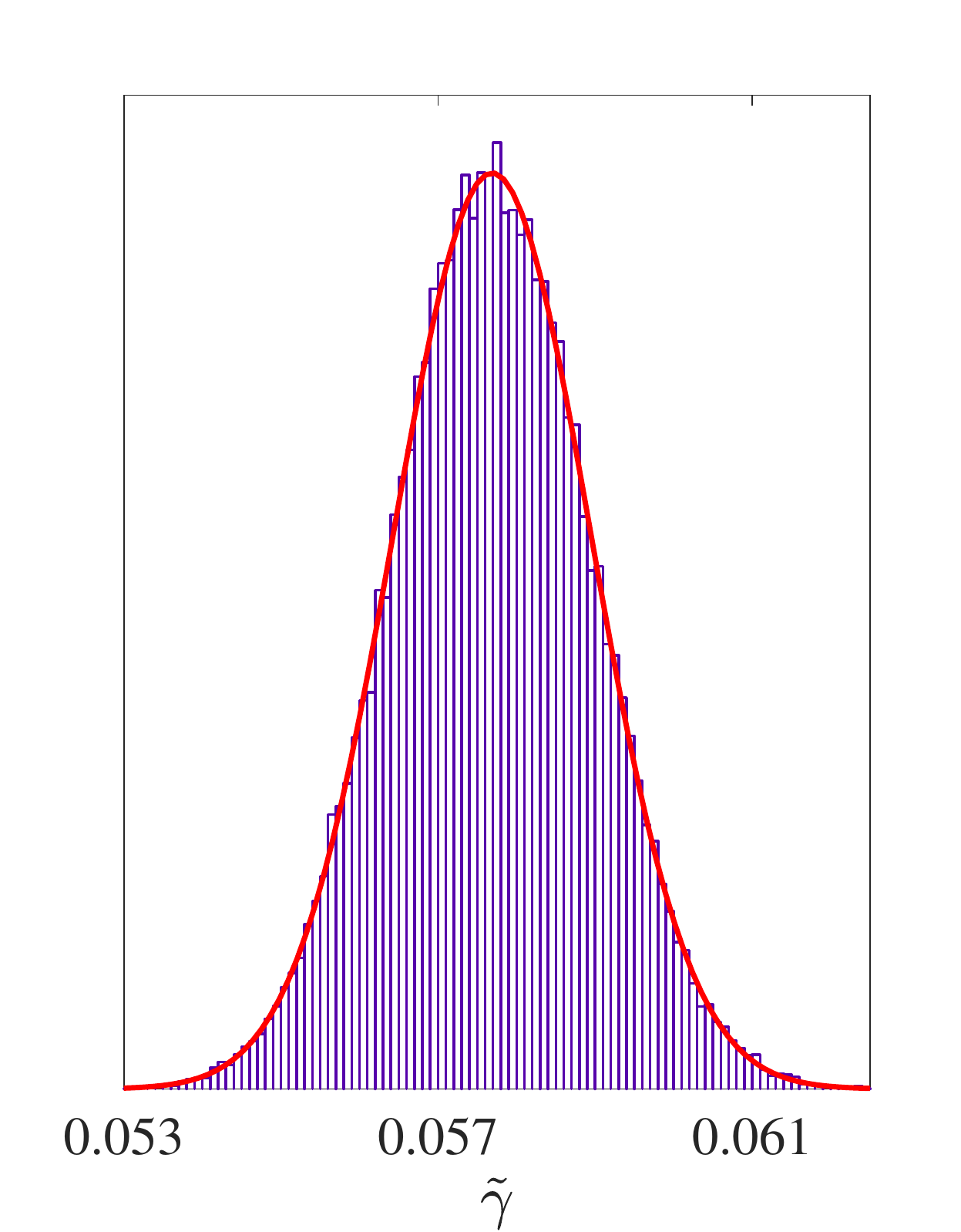}
			\end{minipage}%
		}%
		\subfigure[ U(1) model]{
			\begin{minipage}[t]{0.2\linewidth}
				\centering
				\includegraphics[width=1\linewidth]{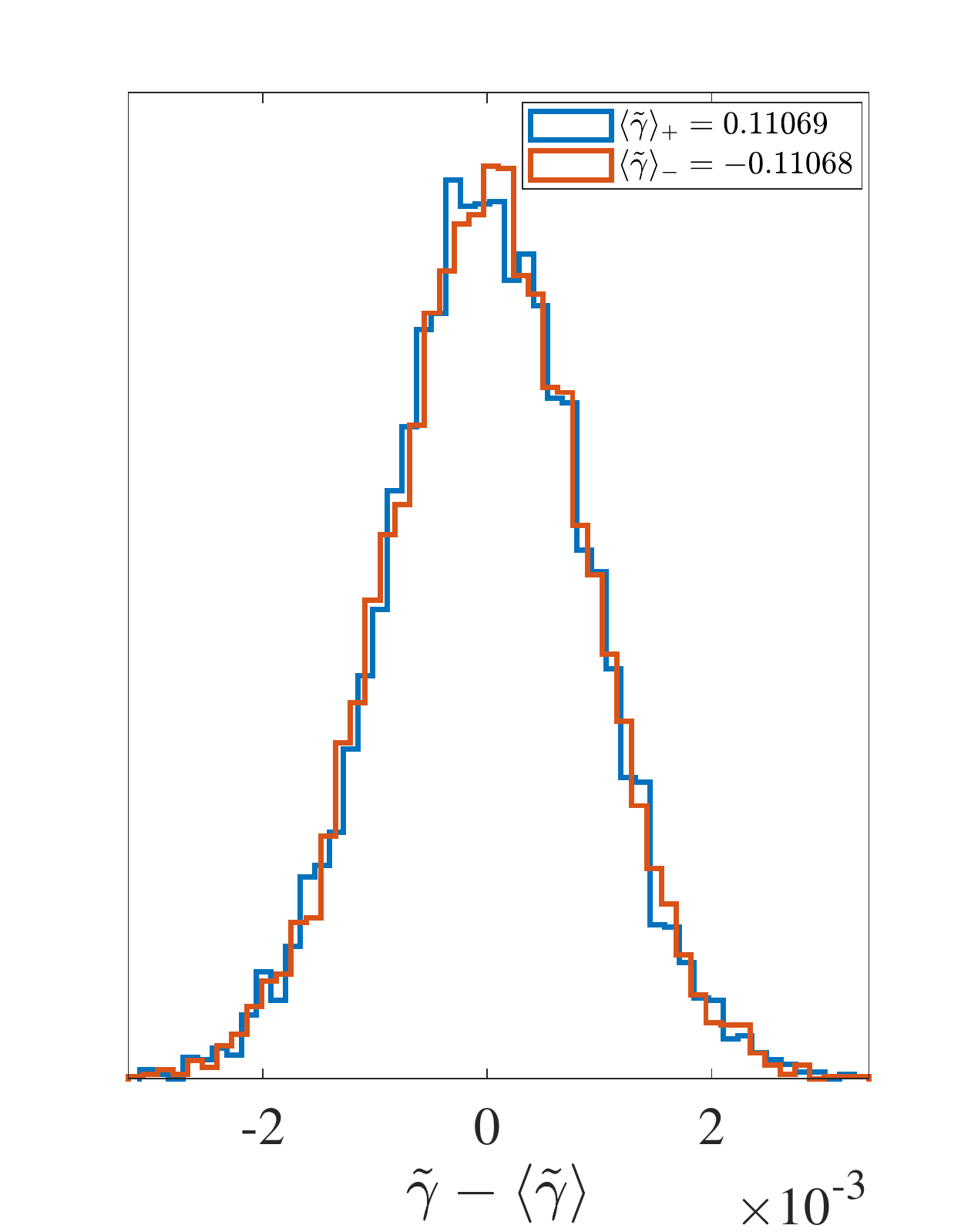}
			\end{minipage}%
		}%
		\subfigure[Hatano-Nelson model]{
			\begin{minipage}[t]{0.2\linewidth}
				\centering
				\includegraphics[width=1\linewidth]{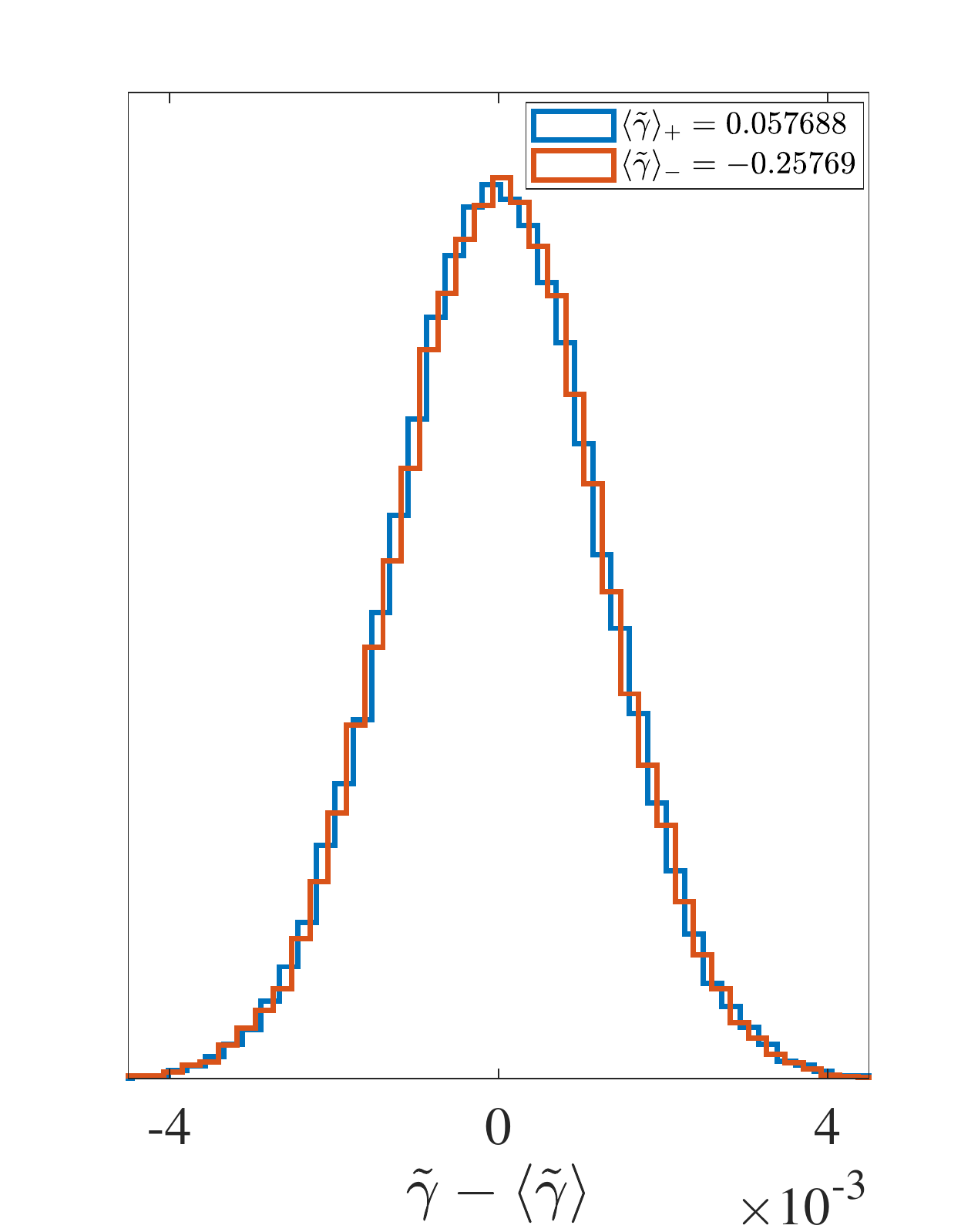}
			\end{minipage}%
		}%
		\caption{Histograms of the Lyapunov exponent $\tilde{\gamma}$ estimated from finite $L_z$.
			For (a)-(c), $\tilde{\gamma}$ stands for the the smallest positive Lyapunov exponent.
			For (d) and (e), $\langle \tilde{\gamma}\rangle_+$ is the smallest positive Lyapunov exponent 
			and $\langle \tilde{\gamma}\rangle_-$ is the largest negative Lyapunov exponent.
			(a) NH Anderson model with $W_r=W_i=6.2$, (b) and (d) NH U(1) model with $W_r=W_i=7$. 
			The data of (a), (b), (d) come from $6400$ realizations of disordered systems with independent random 
			number with $L_x=L_y=L=10$ and $L_z=10^5$ (also $L_z=10^4$ in (a)) at $E=0$.	 
			(c) and (e) 1D Hatano-Nelson model with $g=0.1$ and $W=4$,  $64000$ realizations of disordered systems
			with $L_z=10^5$ at $E=0$. The red curves in (a)-(c) are the Gaussian fittings. 
			}
		\label{gamma}
	\end{figure*}
	
	We consider transmission of particles of a complex-valued energy $E$ through 
	long disordered wire ($L_x,L_y\ll L_z$) with a square cross-section $L_x=L_y=L$. 
	For the long wire, the amplitude of the transmission decays exponentially 
	with an associated decay length called the quasi-1D
	localization length $\lambda$. To calculate $\lambda$, we consider  
	a product of the transfer matrix over the layer from $i_z=1$ to $i_z=L_z$. 
	The product relates the wavefunctions at $i_z=0,1$ and those at $i_z=L_z,L_z+1$;
	\begin{equation}
		M=\prod_{i_z=1}^{L_z} M_{i_z}.	 \label{Mdef}
	\end{equation}
	For the product of random matrices such as $M_{i_z}$, some eigenvalues of the product decay exponentially 
	in $L_z$, while the others grow exponentially in $L_z$. The simultaneous emergence of the eigenvalues 
	with the exponential decay and growth represents a reciprocal symmetry
	nature of the 3D NH systems. The reciprocal symmetry in the NH Anderson model is protected 
	by the transposition symmetry ${\cal H}^\mathrm{T}={\cal H}$ in each disorder realization, while 
	it is not the case in the NH U(1) and Peierls models. Nonetheless, the reciprocal symmetry in the latter 
	two models appears asymptotically in the thermodynamic limit $L_z \rightarrow \infty$, where 
	the transposition symmetry (Hermitian symmetry) is effectively restored after an average 
	over $i_z$ from $i_z=0$ to $i_z=L_z$; the transposition symmetry (Hermitian symmetry)
	exists statistically in the NH U(1) (Peierls) models.
	
	The amplitude part of the complex-valued eigenvalue of $M$ with the slowest exponentially decay 
	(growth) defines the quasi-1D localization length of the system with the 
	long-wire geometry. To extract this length, it is convenient to consider a positive-semi-definite 
	Hermitian matrix $MM^{\dagger}$, and its logarithm,
	\begin{equation}
		\Omega=\ln M M^{\dagger}.	
	\end{equation}
	When ${\cal H}$ is Hermitian, $\Omega$ satisfies the Oseledec's theorem\cite{Furstenberg60,Oseledec68}. The theorem dictates that in the 
	limit of large $L_z$, eigenvalues of $\Omega/L_z$ obtained from the product of random 
	matrices $M_{i_z}$ ($i_z=1,\cdots,L_z$) converges to non-random numbers. 
	Any matrix product of $n\times n$ complex-valued matrices is equivalent to $2n\times 2n$ 
	real-valued matrices, because a product of two complex numbers $z_1=x_1+{\rm i} y_1$ and $z_2 = x_2+{\rm i} y_2$ 
	can be regarded as a matrix product of two real-valued 2 by 2 matrices $x_1 \tau_0 + {\rm i} y_1 \tau_y$ 
	and  $x_2 \tau_0 + {\rm i} y_2 \tau_y$.
	Thereby, $\Omega/L_z$ obtained from the non-Hermitian random 
	${\cal H}$ also satisfies the Oseledec's theorem~\cite{Ruelle79}. According to the theorem, 
	eigenvalues of $\Omega$ decrease (increase) linearly in the large $L_z$. The Lyapunov 
	exponents $\gamma_i$ are defined by the eigenvalues of $\Omega$, $\nu_i$, 
	\begin{equation}
		\gamma_{i} \equiv \frac{\nu_i}{2L_z} 
	\end{equation} 
	for $i=1,2,\cdots,2L_xL_y$. 
	
	In the NH Anderson model, the positive Lyapunov exponents and negative Lyapunov exponents appear in pairs, 
	\begin{align}
		\{\cdots, \gamma_3,\gamma_2, \gamma_1,-\gamma_1,-\gamma_2,-\gamma_3,\cdots\},  \nonumber 
	\end{align}
	with $\cdots>\gamma_2>\gamma_1>0>-\gamma_1>\cdots$ because of the transposition symmetry
	of the 
	transfer matrix;
	\begin{align}
		\sigma_{y}M^\mathrm{T}_{i_z}\sigma_{y}=M^{-1}_{i_z}, \label{eq6}
	\end{align}
	for any $i_z$. The symmetry relates a pair of the two eigenvalues $\{\nu_i, -\nu_i\}$ in any finite $L_z$ 
	through $\sigma_y (MM^{\dagger})^\mathrm{T} \sigma_y = (MM^{\dagger})^{-1}$. In the NH U(1) model, 
	the transpositional symmetry holds between $M_{i_z}$ from ${\cal H}$ and $M_{i_z}$ from ${\cal H}^\mathrm{T}$;
	\begin{align}
		\sigma_{y}M_{i_z}^\mathrm{T}({\cal H})\sigma_{y}=M_{i_z}^{-1}({\cal H}^\mathrm{T}). \label{symplectic-u1}
	\end{align}
	Thus, the positive (negative) Lyapunov exponents from ${\cal H}$ and the negative (positive) Lyapunov
	exponents from ${\cal H}^\mathrm{T}$ are related to each other;
	\begin{align}
		\{\cdots,\gamma_3, \gamma_2, \gamma_1,-\gamma^{\prime}_1,-\gamma^{\prime}_2,-\gamma^{\prime}_3,\cdots\} \nonumber 
	\end{align}
	from ${\cal H}$ with $\cdots>\gamma_2>\gamma_1>0>-\gamma^{\prime}_1>\cdots$ and 
	\begin{align}
		\{\cdots,\gamma^{\prime}_3, \gamma^{\prime}_2, \gamma^{\prime}_1,-\gamma_1,-\gamma_2,-\gamma_3,\cdots\} \nonumber 
	\end{align}
	from ${\cal H}^\mathrm{T}$ with $\cdots>\gamma^{\prime}_2>\gamma^{\prime}_1>0>-\gamma_1>\cdots$. 
	Since ${\cal H}$ and ${\cal H}^\mathrm{T}$ appear with an equal probability in the NH U(1) model, 
	$\gamma_i=\gamma^{\prime}_i$ holds true in the thermodynamic limit ($L_z \rightarrow \infty$). 
	The same holds true in the NH Peierls model, in which ${\cal H}$ and ${\cal H}^{\dagger}$ appear with an equal probability.
	Finally, the reciprocal of the smallest positive Lyapunov exponent $\gamma_1$ is 
	nothing but the quasi-1D localization length
	\begin{equation}
		\lambda=1/\gamma_1 = 2L_z/\nu_1.
	\end{equation}

	To check the stability of the estimate of the Lyapunov exponents in the NH systems by the 
	transfer matrix method, we estimate the Lyapunov exponents 
	 with finite $L_z$ in repeated simulations with the 
	same parameters with an independent stream of random numbers.
	In practice, the Lyapunov exponents are calculated by 
	$QR$ decomposition \cite{Slevin14} instead of by diagonalizing the matrix $\Omega$. 
	Fig.~\ref{gamma} shows distributions of the Lyapunov exponents in 
	the NH Anderson and U(1) models as well as 1D Hatano-Nelson model \cite{Hatano96}. 
	The distributions are always Gaussian, indicating the stability of the evaluations of 
	the Lyapunov exponents from the transfer matrix method. 
	Moreover, the standard deviation of the Gaussian distribution becomes smaller with larger 
	$L_z$ [Fig.~\ref{gamma}(a)], which indicates that Lyapunov exponent will converge 
	to a constant in the large $L_z$ limit. The Lyapunov exponents in the NH Anderson 
	model come in pair protected by symmetry. In the NH U(1) and Peierls models, 
	the smallest positive Lyapunov exponent and the 
	largest negative Lyapunov exponent are consistent with each other 
	in sufficiently large $L_z$ [Fig.~\ref{gamma}(d)]. On the one hand, they are distinct from each other even in the large $L_z$ 
	in the 1D Hatano-Nelson model [Fig.~\ref{gamma}(e)], where both Hermitian symmetry and transposition 
	symmetry are broken even statistically. In the 1D Hatano-Nelson model, rightward-decaying 
	eigenfunctions and leftward-decaying eigenfunctions have different longest decay lengths 
	due to asymmetric hopping terms; $e^{g}$ for rightward hopping term and $e^{-g}$ 
	for leftward hopping term. 
	
	
	\subsubsection{conductance}
	The two-terminal conductance can be also 
	formulated for the NH systems
	in the same framework as the scattering theory in Hermitian systems~\cite{Pendry92,Slevin01}. 
	Thereby, the 3D $L_x \times L_y \times L_z$ cubic lattice is regarded as a scattering 
	object and the two-terminal conductance is defined as a total number of particles that 
	transmit through the scattering object within a unit time and within a unit energy window. 
	To calculate the conductance, we consider that the disordered system with a cross-section of 
	$L_x L_y$ and a length of $L_z$ is attached to two leads. Each lead comprises of 
	$2 L_x L_y$ decoupled 1D wires, half of which have right-moving current flux along $+z$ 
	direction and the other half have left-moving current flux along $-z$. For simplicity, 
	we regard the two leads as the Hermitian system and identify all the $L_x L_y$ 1D wires with 
	the left(right)-moving flux as eigenstates of the following 1D tight-binding 
	model 
	\begin{align}
		{\cal H}_{\rm lead}= \sum_{i_z} c^{\dagger}_{(i_x,i_y,i_z+1)} c_{(i_x,i_y,i_z)} + {\rm h.c.}. 
	\end{align} 
	The left (right)-moving current flux states are regarded as eigenstates at right (left) `Fermi' points  
	of the 1D tight-binding model respectively. Such states are given by $e^{-{\rm i}k_z i_z}$ 
	and $e^{{\rm i}k_z i_z}$ with their 
	(real-valued) eigen-energy $E=2 \cos k_z$ and $k_z \in [-\pi,\pi)$. 
	
	\begin{figure}[tb]
		\includegraphics[width=1\linewidth]{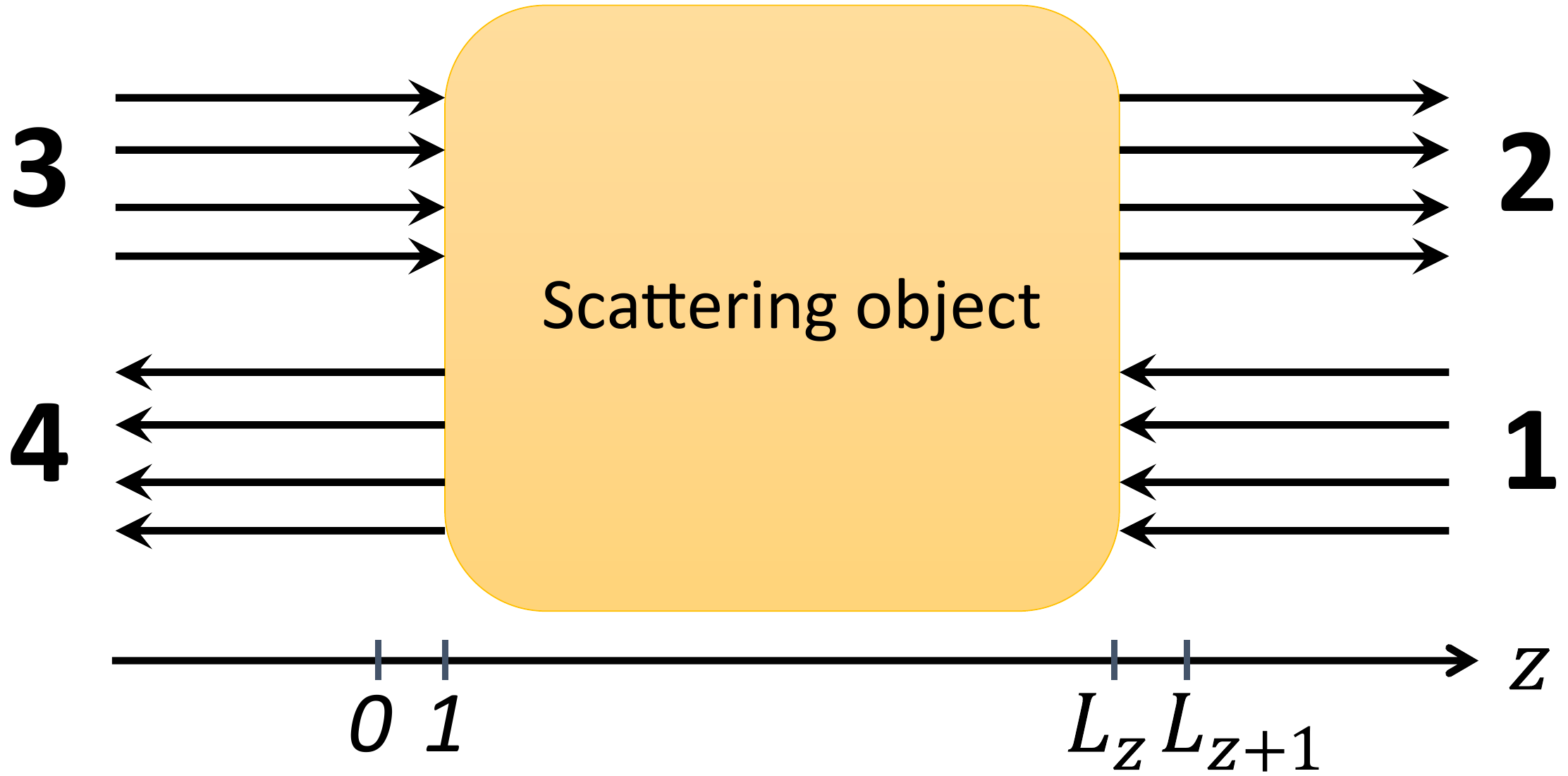}
		\caption{Two-terminal conductance geometry. 1 and 3 represent the incoming current flux;
			2 and 4 represent the outgoing current flux.}
		\label{2-t}
	\end{figure}

	The scattering object is characterized by a scattering matrix $S$ among right-moving current fluxes 
	and left-moving fluxes in the two leads\cite{Datta97};
	\begin{align}
		\left(\begin{array}{c}
			2 \\
			4  \\
		\end{array}\right) = \left(\begin{array}{cc}
			r & t^{\prime}  \\
			t & r^{\prime}  \\
		\end{array}\right) \left(\begin{array}{c}
			1 \\
			3  \\
		\end{array}\right) \equiv S \left(\begin{array}{c}
			1 \\
			3  \\
		\end{array}\right).   \label{s1}
	\end{align}  
	Here $1$ ($2$) and $4$ ($3$) represent $L_x L_y$ vectors for the left (right)-moving fluxes 
	in the right lead and in the left lead, respectively (Fig.~\ref{2-t}). Four 
	$L_xL_y \times L_x L_y$ matrices in $S$, $t$, $t^{\prime}$, $r$ and $r^{\prime}$,  
	are the transmission matrix from right to left, the transmission matrix from left to right, the reflection 
	matrix in the right lead, and the reflection matrix in the left lead, respectively. Due to the reciprocal symmetry, 
	the scattering matrix is symmetric ($S^\mathrm{T}=S$) in the NH Anderson model, while they are not in 
	the NH U(1) and NH Peierls models.

	In the NH U(1) and Peierls models, the transposition of the Hamiltonian changes the sign of 
	the external magnetic fluxes $\{\Phi\}$, where $\{\Phi\}$ indicates the position dependent
	magnetic fluxes in the U(1) model whereas it is constant in the Peierls model.
	Thereby, the leftward conductance and rightward conductance 
	defined below do not coincide with each other in general for a nonzero $\{\Phi\}$.
	\begin{align}
		g_L &\equiv  {\rm Tr}[tt^{\dagger}], \nonumber \\
		g_R &\equiv  {\rm Tr}[t^{\prime}{t^{\prime}}^{\dagger}]. \nonumber 
	\end{align}
	Nonetheless, $S^\mathrm{T}({\cal H}_{\{\Phi\}})= 
	S({\cal H}^\mathrm{T}_{\{\Phi\}})=S({\cal H}_{\{-\Phi\}})$ indicates that leftward (rightward) 
	conductance with the magnetic fluxes ${\{\Phi\}}$ is the same as the rightward (leftward) conductance with 
	the flux ${\{-\Phi\}}$ in the NH U(1)/Peierls models;
	\begin{align}
		g_{R/L}({\{\Phi\}}) = g_{L/R}({-\{\Phi\}}). \label{NH-Peierls}
	\end{align}

	To calculate the transmission matrix $t$ by the transfer matrix  method, we rewrite the scattering 
	matrix into a matrix $T$.
	$T$ connects the left lead to the right lead;
	\begin{align}
		\left(\begin{array}{c}
			1 \\
			2  \\
		\end{array}\right) = \left(\begin{array}{cc}
			t^{-1} & -t^{-1} r^{\prime}  \\
			r t^{-1} & -rt^{-1}r^{\prime}+t^{\prime}  \\
		\end{array}\right) \left(\begin{array}{c}
			4 \\
			3  \\
		\end{array}\right) \equiv T \left(\begin{array}{c}
			4 \\
			3  \\
		\end{array}\right),   \label{whatisT}
	\end{align}   
	The product of the transfer matrix, $M$, relates wavefunction amplitudes 
	of ${\cal H}$ in the left and the right leads. 
	Thereby, by the use of $M$, the eigenstates of ${\cal H}_{{\rm lead}}$ (the left lead)  
	with the two opposite current flux can be generalized into the eigenstates of 
	${\cal H}+{\cal H}_{\rm lead}$. 
	The two eigenstates thus obtained take the following forms 
	at the beginning of the right lead ($i_z=L_z+1,L_z$);  
	\begin{align}
		\left(\begin{array}{c}
			\psi_{L_z+1} \\
			\psi_{L_z}  \\
		\end{array}\right) &= M \frac{1}{\sqrt{2}}\left(\begin{array}{c}
			e^{{\rm i}k_z} \\
			1  \\
		\end{array}\right) \otimes {\bm e}_i, \nonumber \\
		\left(\begin{array}{c}
			\psi_{L_z+1} \\
			\psi_{L_z}  \\
		\end{array}\right) &= M \frac{1}{\sqrt{2}}\left(\begin{array}{c}
			1 \\
			e^{{\rm i}k_z}  \\
		\end{array}\right) \otimes {\bm e}_i, \nonumber 
	\end{align}
	with $k_z>0$ and $i=1,2,\cdots, n$, $n$ being $L_x L_y$. Here we note that 
	$V_{i_z+1,i_z}$ in Eq.~(\ref{eq2}) is an $n$-dimensional 
	unit matrix for the NH Anderson and U(1) models  and ${\bm e}_i$ is an 
	$n$-dimensional unit vector associated with $x$ and $y$ coordinates.
	In the following,  we will omit the $x$-$y$ coordinate degree of freedom unless dictated explicitly. 
	
	The matrix $T$ is given by an inner product between the two eigenstates of 
	${\cal H}+{\cal H}_{\rm lead}$ and the two eigenstates of the right-lead Hamiltonian; 
	\begin{align}
		T \equiv \frac{1}{2} \left(\begin{array}{cc}
			1 & e^{-{\rm i}k_z}\\
			e^{-{\rm i}k_z}  & 1 \\
		\end{array}\right) \Sigma M  \left(\begin{array}{cc}
			1 & e^{{\rm i}k_z}\\
			e^{{\rm i}k_z}  & 1 \\
		\end{array}\right).  
	\end{align}
	The inner products are taken at $i_z=L_z+1, L_z$ with a matrix $\Sigma$. 
	When all the $H_{i_z}$ in $M$ are set to zero, 
	the scattering object becomes identical to the two leads. Thereby, the $T$ matrix at $E=2\cos k_z$ 
	must be proportional to a diagonal matrix with diagonal elements of the unit modulus. 
	This requirement defines the inner product with a proper normalization\cite{Pendry92,Ando91};
	\begin{align}
		T &\equiv \frac{1}{2 \sin k_z} \left(\begin{array}{cc}
			1 & e^{-{\rm i}k_z}\\
			e^{-{\rm i}k_z}  & 1 \\
		\end{array}\right) \sigma_y M  \left(\begin{array}{cc}
			1 & e^{{\rm i}k_z}\\
			e^{{\rm i}k_z}  & 1 \\
		\end{array}\right). \label{T-def}
	\end{align} 
	For the Hermitian system, the symplectic nature of the transfer matrix guarantees the 
	unitarity of the scattering matrix\cite{Datta97}. Namely, $\sigma_y M^{\dagger} \sigma_y = M^{-1}$ 
	leads to $\sigma_z T^{\dagger}\sigma_z = T^{-1}$ and $S^{\dagger} = S^{-1}$.
	
	In the NH Anderson model, the transfer matrix only has a pseudo-symplectic nature, 
	Eq.~(\ref{eq6}). The pseudo-symplecticy of the transfer matrix imposes 
	the pseudo-symplecticy on $T$ and the symmetric nature on the $S$ matrix 
	respectively; 
	\begin{align}
		\sigma_y T^\mathrm{T} \sigma_y = T^{-1}, \!\ S^\mathrm{T} = S. \label{pseudo-symmetric}
	\end{align}
	The symmetric $S$ matrix leads to the reciprocal symmetry in the conductance; 
	$g_R=g_L$. 
	In the NH U(1) model, the pseudo-symplecticy holds true only 
	between ${\cal H}$ and ${\cal H}^\mathrm{T}$. So does the symmetric nature of the $S$ matrix;
	\begin{align}
		S^\mathrm{T}({\cal H}) = S({\cal H}^\mathrm{T}). \label{ps-nh-u1}
	\end{align}
	Since ${\cal H}$ and ${\cal H}^\mathrm{T}$ appear with equal probability in the NH U(1) model, 
	eq.~(\ref{ps-nh-u1}) leads to the reciprocal symmetry in the averaged conductance in 
	the NH U(1) model; $\langle g_R \rangle = \langle g_L \rangle$. Thereby, for the NH Anderson 
	and U(1) models, we have only to define the two-terminal conductance by one of the two transmission 
	matrices;     
	\begin{align}
		g\equiv \langle {\rm Tr}[tt^{\dagger}]\rangle. \label{two-terminal-conductance}
	\end{align}
	In the NH Peierls model where the magnetic field is uniform, $g_R$ and $g_L$ are different. 
	

	Note that in the three NH models, the non-Hermiticity is introduced only by the on-site 
	NH disorder potential, $w_{\bm i}^r$ and $w_{\bm i}^i$, which range in $[-W_r/2,W_r/2]$ and $[-W_i/2,W_i/2]$, 
	respectively. Thereby, the Hermiticity is recovered statistically; ${\cal H}^{\dagger}$ and ${\cal H}$ 
	realize with equal probability in an ensemble of many disorder realizations. 
	Nonetheless, $S^{\dagger}({\cal H}) = S^{-1}({\cal H}^{\dagger})$ does not guarantee the 
	unitarity of the scattering matrix in any way, e.g. even after the disorder average, 
	\begin{align}
		\langle r^{\dagger} r \rangle + \langle t^{\dagger} t \rangle \ne {\mathbb 1}_{L_xL_y \times L_xL_y}. \label{no-use}
	\end{align}  
	On the contrary, $S^{\dagger}({\cal H}) = S^{-1}({\cal H}^{\dagger})$  gives 
	\begin{align}
		r^{\dagger}({\cal H}) r ({\cal H}^{\dagger}) + t^{\dagger}({\cal H}) t ({\cal H}^{\dagger}) = 
		{\mathbb 1}_{L_xL_y \times L_xL_y}. \label{no-use-2}
	\end{align}
	This suggests that the conductance $g$ in the NH systems is {\it not} bounded by the total 
	number of the transmission channels, i.e.,  $n=L_xL_y$. This is in some sense physically reasonable 
	because the transmission amplitude in the NH systems can be amplified by the on-site 
	NH disorder potentials, when particles going through the scattering object.

    \subsubsection{Lyapunov exponent and scattering matrix in non-Hermitian systems 
    	with pseudo-Hermiticity}\label{appendix_Pseudo}
    Transfer matrix method has been 
    extensively used in ${\cal PT}$-symmetric 
    non-Hermitian optical systems~\cite{Artoni05,Reddy13,Ramezani14,Mostafazadeh09,Basiri15,Loran16,Mullers18,Konotop16,Simon19,Longhi10,Schomerus10,Chong11,Schomerus11,
    	Yoo11,Ge12,Schomerus13} as well as pseudo-Hermitian magnon
    systems~\cite{Xu16}. The ${\cal PT}$-symmetric systems can be 
    regarded as non-Hermitian system with 
    pseudo-Hermiticity~\cite{Mostafazadeh02a,Mostafazadeh02b}. In this section,  
    we summarize reciprocal symmetry of the Lyapunov 
    exponents and symmetry properties of the scattering matrix in 
    non-Hermitian disordered systems with pseudo-Hermiticity. 
    The symmetry properties of the scattering 
    matrix have been previously  
    discussed in the ${\cal PT}$ symmetric 
    optical systems in the context of perfect 
    coherent absorber 
    \cite{Longhi10,Schomerus10,Chong11,Schomerus11,
    	Yoo11,Ge12,Schomerus13}, and 
    1D NH class AI (Hatano-Nelson model)~\cite{Longhi15}. 
    
    Hamiltonian $\mathbb{H}$ with the pseudo-Hermiticity,   
    \begin{align}
    	\eta {\mathbb{H}} \eta^{-1} = {\mathbb{H}}^{\dagger}, \label{ph}
    \end{align} 
    is introduced in an eigenvalue problem on a 3D cubic lattice, 
    \begin{align}
    	\mathbb{H} \Psi = E \Psi.  
    \end{align}
    The argument can be easily generalized into other lattices. 
    Here $\eta$ is Hermitian and unitary matrix;  
    $\eta^{\dagger}\eta=\eta\eta^{\dagger}=1$ and  
    $\eta^{\dagger}=\eta$. We allow $\mathbb{H}$ to have
    internal degrees of freedom, 
    such as spin, sublattice, and particle-hole degrees of freedom. For example, 
    generalized eigenvalue problems for free quasi-particle 
    boson systems are equivalent to 
    diagonalizing pseudo-Hermitian Hamiltonian~\cite{Shindou13a,Shindou13b,Kawabata19,Okamoto20}. Thereby, 
    the internal degree of freedom is particle-hole degree of freedom 
    of the boson, $\Psi$ is a Nambu vector subtended by 
    both boson creation and annihilation operators and 
    $\eta$ is given by a diagonal matrix that 
    takes $+1/-1$ for the 
    creation/annihilation operators~\cite{Shindou13a,Shindou13b}. 
    As above, 3D $L_x\times L_y\times L_z$ cubic lattice is regarded as 
    the multiple layer structure of the two-dimensional (2D) slices. 
    Without loss of generality, we can assume that $\mathbb{H}$ has 
    hoppings only between the nearest neighboring 2D slices, %
    \begin{align}
    	\mathbb{H} = \left(\begin{array}{cccc}
    		\ddots & V_{i_z-1,i_z} & 0 &  \\
    		V_{i_z,i_z-1} & H_{i_z} & V_{i_z,i_z+1} & 0 \\
    		0 & V_{i_z+1,i_z} & H_{i_z+1} & V_{i_z+1,i_z+2} \\
    		& 0 & V_{i_z+2,i_z+1} & \ddots \\
    	\end{array}\right). 
    \end{align}
    $H_{i_z}$ includes on-site disorder potentials 
    and hopping integrals within the 2D slice at $i_z$. 
    $V_{i,j}$ are hopping integrals between 
    2D slice at $i$ and 2D slice at $j$, which we assume to be 
    any invertible $L_xL_y \times L_xL_y$ matrix. In the presence 
    of the on-site disorder in $H_{i_z}$, a Hermitian 
    matrix $\eta$ that satisfies Eq.~(\ref{ph}) must be diagonal 
    with respect to the site index ${\bm i}=(i_x,i_y,i_z)$. 
    Accordingly, the transfer 
    matrix in Eq.~(\ref{eqn2a}) has the following symmetry,
    \begin{align}
    	\eta \sigma_y M^{\dagger}_{i_z}(E) \sigma_y \eta^{-1} = {M_{i_z}(E^{*})}^{-1} 
    	\label{sup1}
    \end{align}
    with complex-valued eigen-energy $E$. 
    This results in the reciprocal symmetry between the Lyapunov 
    exponents at $E$ and $E^*$;
    \begin{align}
    	\{\cdots, \gamma_2(E), \gamma_1(E),-\gamma^{\prime}_1(E),   -\gamma^{\prime}_2(E),\cdots\} \nonumber 
    \end{align}
    with $\gamma_i(E^*)=\gamma^{\prime}_i(E)$, 
    $\cdots>\gamma_{2}>\gamma_1>0>-\gamma^{\prime}_1>-\gamma^{\prime}_2>\cdots$. 
    For real-valued $E$, $\gamma^{\prime}_i(E)=\gamma_i(E)$. Based on this reciprocal 
    relation, Ref.~\cite{Xu16} previously 
    studied quantum magnon Hall plateau transition, 
    clarifying that the universality class of 
    the plateau transition in the pseudo-Hermitian systems  
    belongs to the same universality class as 
    the Hermitian quantum 
    Hall plateau transition. Eq.~(\ref{sup1}) also dictates 
    that the scattering matrix defined in 
    Eqs.~(\ref{s1}), (\ref{whatisT}), and (\ref{T-def}) has  
    unitarity-like relation between $E$ and $E^*$;  
    $S^{\dagger}(E) \eta S(E^*) = \eta$. 
    
    The pseudo-Hermiticity allows time-independent inner product 
    between two wavefunctions in the Schrodinger picture; 
    $\partial_t (\langle \Psi(t)|\eta |\Psi(t)\rangle) =0$ 
    with $\mathbb{H}|\Psi(t)\rangle= i\partial_t |\Psi(t)\rangle$. 
    Such an inner product represents conserved (local) 
    physical quantities of underlying physical systems. 
    In the example of the generalized 
    eigenvalue problems for quasi-particle  
    boson systems, the inner product 
    corresponds to an energy density carried by the  
    bosons~\cite{Haldane08,Raghu08,Xu16,Okamoto20}. 
    Thus, the transfer matrix method in such pseudo-Hermitian systems should be 
    reformulated in such a way that the conservation 
    rule becomes explicit in the transport properties. 
    To this end, it is more natural to introduce the 
    transfer 
    matrix as~\cite{Xu16} 
    \begin{align}
    	\tilde{M}_{i_z}=\begin{pmatrix} \tilde{V}_{i_z,i_z+1}^{-1}(E\eta -\tilde{H}_{i_z}) &-\tilde{V}_{i_z,i_z+1}^{-1}\\ \tilde{V}_{i_z+1,i_z}& 0\end{pmatrix}, \label{eq:psuedoHermiticity}
    \end{align}
    with 
    \begin{align}
    	\tilde{\mathbb{H}} \equiv \eta \mathbb{H} 
    	\equiv \left(\begin{array}{cccc}
    		\ddots & \tilde{V}_{i_z-1,i_z} &  & \\
    		\tilde{V}_{i_z,i_z-1} & \tilde{H}_{i_z} & \tilde{V}_{i_z,i_z+1} &  \\
    		& \tilde{V}_{i_z+1,i_z} & \tilde{H}_{i_z+1} & \tilde{V}_{i_z+1,i_z+2} \\
    		&  & \tilde{V}_{i_z+2,i_z+1}  & \ddots \\
    	\end{array}\right), 
    \end{align} 
    and calculate $T$-matrix and scattering matrix according to 
    Eqs.~(\ref{s1},\ref{whatisT},\ref{T-def}) with $M$ replaced by
    $\tilde{M} \equiv \prod_{i_z} \tilde{M}_{i_z}$. Such scattering 
    matrix respects the unitarity relation, 
    $S^{\dagger}(E) S(E^{*})=\mathbb{1}$, and therefore 
    the conductance for the real-valued $E$ is compatible 
    with the conservation principle. In the example of the free 
    quasi-particle boson systems, 
    the conductance thus calculated 
    is thermal (energy) conductance carried
    by the quasi-particle bosons~\cite{Xu16}. Note also 
    that $\tilde{M}$ gives the same sets of the 
    Lyapunov exponents as $M$, since  $\tilde{M}$ 
    and $M$ are related to each other by a unitary transformation. 
    
    	Before concluding this subsection, we like to emphasize that the models of 
    	the NH class A and NH class AI$^\dagger$ studied in this paper
    	do not have ${\cal PT}$-symmetry, but they still have symmetry relations 
    	such as
Eqs.~(\ref{eq6}), (\ref{symplectic-u1}), (\ref{NH-Peierls}), (\ref{pseudo-symmetric}), (\ref{ps-nh-u1}), and (\ref{no-use-2}): see a summary in Table II.

	\subsection{polynomial fitting}
	\label{subsec:polynomiaFitting}
	\begin{table*}[tb]
		\centering
		\setlength{\tabcolsep}{1.1mm}
		\caption{Polynomial fitting results for normalized localization length around 
			the Anderson transition point in the presence of the NH disorder with disorder strength 
			$W_r$ and $W_i$. The goodness of fit (GOF), critical disorder $W_{c}$, critical 
			exponent $\nu$, the scaling dimension of the least irrelevant scaling variable $-y$, 
			and the critical normalized localization length $\Lambda_c$ are 
			shown for various system sizes and for different 
			orders of the Taylor expansion: $(m_1,n_1,m_2,n_2)$. 
			The square bracket is the 95\% confidence interval.}
		\begin{tabular}{ccccccccccc}
			&&&&&&&&&&\\
			\multicolumn{11}{l}{Anderson model at $E=0$}\\
			\hline
			disorder&	$L$&	$m_1$& $n_1$ & $m_2$ & $n_2$ & GOF & $W_c$ & $\nu$ & $y$&$\Lambda_c$   \\
			\hline
			\multirow{3}*{$W_r=W_i$}
			&6-24&3&3&0&1&0.10&6.3868[6.3861, 6.3874]&1.190[1.187, 1.193]&2.33[2.22, 2.48]&0.8359[0.8350, 0.8369]\\
			&8-24&3&3&0&1&0.16&6.3858[6.3851, 6.3864]&1.193[1.189, 1.197]&2.57[2.43, 2.72]&0.8375[0.8365, 0.8384]\\
			&10-24&3&3&0&1&0.13&6.3862[6.3852, 6.3872]&1.192[1.185, 1.198]&2.67[2.44, 2.91]&0.8369[0.8353, 0.8385]\\
			\hline
			$W_i=5$
			&6-16& 3&3&0&1&0.12&7.839[7.837, 7.842]&1.194[1.189, 1.199]&2.38[2.25, 2.52]&0.8368[0.8356, 0.8380]\\
			$W_r=5$
			&6-16&3&3&0&1&0.19&7.841[7.838, 7.843]&1.190[1.185, 1.196]&2.37[2.24, 2.57]&0.8361[0.8348, 0.8376]\\
			\hline
			$W_i=0.1$
			&10-20&3&3&1&1&0.18&14.435[14.414, 14.462]&0.914[0.843, 1.026]&0.99[0.84, 1.25]&0.8442[0.8246, 0.8600]\\
			\hline
			\multirow{2}*{$W_r=0$}
			&4-20&2&3&0&1&0.90&16.543[16.536, 16.550]&1.574[1.567, 1.581]&2.67[2.42, 2.92]&0.5753[0.5742, 0.5763]\\
			&4-20&3&3&0&1&0.90&16.543[16.535, 16.550]&1.569[1.558, 1.578]&2.70[2.46, 2.97]&0.5754[0.5743, 0.5764]\\
			\hline
			&&&&&&&&&&\\
			\multicolumn{11}{l}{Anderson model at $E={\rm i}$}\\
			\hline
			disorder&	$L$&	$m_1$& $n_1$ & $m_2$ & $n_2$ & GOF & $W_c$ & $\nu$ & $y$&$\Lambda_c$   \\
			\hline
			$W_r=W_i$
			&6-16&3&3&0&1&0.12&6.018[6.015, 6.022]&1.183[1.172, 1.192]&2.46[1.96, 3.13]&0.8311[0.8278, 0.8341]\\
			\hline
			&&&&&&&&&&\\
			\multicolumn{11}{l}{Anderson model at $E=2$}\\
			\hline
			disorder&	$L$&	$m_1$& $n_1$ & $m_2$ & $n_2$ & GOF & $W_c$ & $\nu$ & $y$&$\Lambda_c$   \\
			\hline
			\multirow{3}*{$W_r=0$}
			&8-20& 2&3&0&1&0.15&11.107[11.099, 11.112]&1.202[1.193, 1.209]&3.39[2.14, 5.02]&0.836[0.833, 0.842]\\
			&8-20&3&3&0&1&0.12&11.108[11.104, 11.111]& 1.208[1.199, 1.215]&3.44[2.62, 4.51]&0.836[0.834, 0.838]\\
			&10-20&3&3&0&1&0.12&11.110[11.106, 11.114]&1.205[1.191, 1.216]&2.90[2.38, 4.03]&0.834[0.832, 0.837]\\
			\hline
			&&&&&&&&&&\\
			\multicolumn{11}{l}{U(1) model at $E=0$}\\
			\hline
			disorder&	$L$&$m_1$& $n_1$ & $m_2$ & $n_2$ & GOF & $W_c$ & $\nu$ & $y$ & $\Lambda_c$ \\
			\hline
			\multirow{3}*{$W_r=W_i$}
			&8-24&1&4&0&1&0.13& 7.322[7.313, 7.330]&1.050[1.010, 1.093]&0.28[0.16, 0.46]&0.434[0.313, 0.514]\\
			&10-24&1&4&0&1&0.11&7.321[7.309, 7.331]&1.041[0.943, 1.096]&0.32[0.15, 0.67]&0.455[0.305, 0.552]\\
			&12-24&3&3&0&1&0.12&7.299[7.297, 7.301]&1.003[0.985, 1.018]&1.59[1.36, 1.89]&0.598[0.593, 0.605]\\
			\hline
			&&&&&&&&&&\\
			\multicolumn{11}{l}{Peierls model with $\Phi=1/4$ at $E=0$}\\
			\hline
			disorder&	$L$&$m_1$& $n_1$ & $m_2$ & $n_2$ & GOF & $W_c$ & $\nu$ & $y$ & $\Lambda_c$ \\
			\hline
			\multirow{3}*{$W_r=W_i$}
			&8-20& 1&4&0&1&0.11&7.077[7.066, 7.087]&1.013[0.932, 1.058]&0.33[0.19, 0.51]&0.475[0.359, 0.545]\\ 
			&10-20&2&4&0&1&0.25&7.047[7.043, 7.050]&1.019[1.011, 1.027]&1.58[1.28, 1.98]&0.628[0.619, 0.637]\\
			&10-20&3&4&0&1&0.23&7.048[7.042, 7.055]&1.020[1.011, 1.033]&1.53[1.01, 2.10]&0.627[0.604, 0.640]\\
			\hline
		\end{tabular}
		\label{table}
	\end{table*}
	
	
	Numerical simulations in the next section show that in the NH systems, 
	the normalized localization length $\Lambda\equiv \lambda/L$ and $g$,  
	and the level spacing ratio 
	(calculated previously in refs.\cite{Huang20,Luo21}) exhibit scale-invariant behaviors at the same critical disorder strength. 
	The scale invariant point is nothing but the Anderson transition point in these NH 
	systems. Quantum criticality of the Anderson transition is universally characterized by critical 
	exponents that depend only on the spatial dimension and the symmetry of the disordered systems. 
	Here, the NH Anderson model belongs to the symmetry class AI$^{\dagger}$, while the NH U(1) 
	and NH Peierls models belong to the symmetry class A. 
	In this subsection, we first review a finite size scaling (FSS) analysis used in this paper. 
	In the next two sections, we present results of the FSS analyses of $\Lambda$ together 
	with new evaluations of the critical exponents in the NH Anderson, U(1) and Peierls models. 
	The critical exponent of the NH U(1) model and that of the Peierls model coincide
	with each other very well, being consistent with the symmetry of these two models. 
	The critical exponent in the NH Anderson model turns out to be clearly distinct from 
	the critical exponent in these two class A models.  
	
	Criticality of any second-order phase transition is determined by a scaling property around 
	a saddle point fixed point for a certain low-energy effective theory. The saddle-point fixed 
	point has only one relevant scaling variable, $\phi_1$, with positive scaling dimension, $1/\nu$. 
	All the other scaling variables, $\phi_2$, $\phi_3$, $\cdots$, are irrelevant with negative 
	scaling dimensions, $-y$, $-y^{\prime}$, $\cdots$. A standard scaling argument 
	dictates that around the transition point, any dimensionless physical 
	quantity $\Gamma$ must be given by a universal function of the scaling variables;  
	\begin{equation}
		\Gamma (W,L)  = F(\phi_1,\phi_2,\phi_3,\cdots). \label{scaling}
	\end{equation}
	The quantity in the left hand side depends on the system size $L$ and a system parameter $W$. 
	Each scaling variable in the right hand side depends on $L$ in power of its scaling dimension;     
	\begin{align}
		\phi_1 &\equiv u_1(w) L^{1/\nu}, \nonumber	\\
		\phi_2 &\equiv u_2(w) L^{-y}, \nonumber \\
		\phi_3& \equiv u_3(w) L^{-y^{\prime}}, \nonumber \\
		&\cdots. \nonumber
	\end{align} 
	Here $1/\nu \!\ (>0)$ is the scaling dimension of the relevant scaling variable and $-y$ is 
	the scaling dimension of the least irrelevant scaling variable; $\cdots<-y^{\prime}<-y \!\ (<0)$. 
	In this paper, we use the disorder strength as the system parameter $W$, and 
	$w$ is a (normalized) distance of $W$ from a critical disorder strength $W_c$; 
	$w\equiv (W-W_c)/W_c$. $u_i(w)$ ($i=1,2,\cdots$) is a function of $W$ with 
	$u_1(w=0)=0$ and $u_i(w=0) \ne 0$ ($i=2,3,\cdots$). When $W$ is sufficiently close 
	to $W_c$, $u_{i}(w)$ can be expanded in power of small $w$\cite{Slevin99,Slevin14};  
	\begin{equation}
		u_{i}(w) \equiv \sum^{m_i}_{j=0} b_{i,j} w^{j}	 \label{uiw}
	\end{equation}
	with $i=1,2,\cdots$, $b_{1,0}=0$, and $b_{i,0} \ne 0$ ($i=2,3,\cdots$). 
	When $w$ is tiny and the system size $L$ is large enough, all the scaling variables 
	are small. For the quantity of such small $w$ and large $L$, the universal function in 
	the right hand side can be expanded in powers of its small arguments\cite{Binder81}. Empirically, we 
	keep only the relevant scaling variable $\phi_1$ and the least irrelevant scaling variable 
	$\phi_2$, while assuming the other irrelevant scaling variables to be zero, $\phi_3=\cdots=0$;     
	\begin{equation}
		F = \sum^{n_1}_{j_1=0} \sum^{n_2}_{j_2=0} a_{j_1,j_2} \phi^{j_1}_1 \phi^{j_2}_2. \label{Fexp}
	\end{equation}
	The assumption is a posteriori  justified with non-small $|y|$ obtained from the fitting (see below).  
	Given $(m_1,n_1,m_2,n_2)$ in Eqs.~(\ref{uiw}),(\ref{Fexp}), $F$ is a  
	finite-order of polynomial of $w \equiv (W-W_c)/W_c$. Numerical data of $\Gamma$ 
	for different $L$ and $W$ are fitted by the polynomial  
	with fitting parameters $W_c$, $\nu$, $y$, $a_{i,j}$ and $b_{i,j}$.
	We minimize $\chi^2$ in terms of the fitting parameters,    
	\begin{equation}
		\chi^2 \equiv \sum^{N_D}_{k=1} \frac{(\Gamma_{k}-F_{k})^2}{\sigma^2_{\Gamma_{k}}}.	\label{key}
	\end{equation}
	Here $k$ counts the data points ($k=1,\cdots, N_D$), and each data point 
	is specified by $L$ and $W$; $k=(L,W)$. $\Gamma_{k}$ and $\sigma_{\Gamma_{k}}$ 
	are a mean value of $\Gamma$ and its standard deviation at $k=(L,W)$, respectively, 
	while $F_{k}$ is fitting value from the polynomial $F$ at $k=(L,W)$. 
	$F_k$ depends 
	on the fitting parameters and $\chi^2$ is minimized in terms of them. 
	The minimization is carried out for several different choices of $(m_1,n_1,m_2,n_2)$. 
	Table~\ref{table} shows fitting results with the goodness of fit greater than 0.1.
	The 95$\%$ confidence intervals for the fitting results are determined by 1000 sets of 
	$N_D$ synthetic data points,
	which are statistically generated from the fitting value $F_k$ with
	the same standard deviation of $\Gamma_k$ at each point $k$.
	
	
	\section{non-Hermitian Anderson model}
	\label{sec:NHAM}
	A phase diagram of the NH Anderson model at $E=0$ is determined in a two-dimensional 
	plane subtended by $W_r$ and $W_i$; Fig.~\ref{phasediagram}(a). The phase boundary between localized and 
	delocalized phases is determined by the scale-invariant point of the normalized localization 
	length $\Lambda_z\equiv\lambda/L$ with $L_x=L_y=L$. 
	$\Lambda_z$ increases with the  size $L$ in the delocalized phase, while 
	decreases in the localized phase. The phase transition is nothing but the Anderson transition 
	in the Hermitian limit ($W_i=0$), where the critical disorder is consistent with a literature 
	value; $W_{r,c}\approx16.5$~\cite{Slevin18}. The phase diagram at $E=0$ is symmetric 
	with respect to an exchange of $W_r$ and $W_i$. 
	
	There are two significant features in the phase diagram. Firstly, the phase boundary 
	of $\{W^2_r+W^2_i\}^{1/2}$ bends in toward the smaller disorder value than the critical disorder 
	strength in the Hermitian case. This means that states are more easily localized by 
	the NH on-site disorder than by the Hermitian on-site disorder with the same modulus. 
	This tendency in 3D Anderson model is consistent with the 2D Anderson model, which 
	shows no localization-delocalization transition with the on-site NH disorder \cite{Huang20}. 
	The second significant feature is that the phase boundary is symmetric along the 
	line of $W_r=W_i$, indicating that the role of $W_r$ and that of 
	$W_i$ are identical in the Anderson localization at $E=0$. In fact, the critical point along 
	an axis of $W_r=0$ as well as along an axis of $W_i=0$ belongs to the 3D Hermitian 
	orthogonal class; the critical exponent evaluated 
	at $W_{i,c}\approx 16.5$ along the axis of $W_r=0$ is consistent with the critical exponent 
	in the orthogonal class in the Hermitian case, $\nu\approx 1.57$ (TABLE \ref{table}).

The symmetric nature of the phase diagram in $W_r-W_i$-plane at $E=0$ is due to
the bipartite lattice structure and the choice of the onsite potential the real and imaginary parts of 
which obey the same distribution. To see this, note that a diagonalization of 
these disordered Hamiltonians at  $E=0$ is clearly inclusive of solving transfer matrices 
in favor for the Lyapunov exponents at $E=0$ for these models;
	\begin{align}
		[\mathbb{H}] \Psi = E \Psi = 0.  \label{E=0}
	\end{align}
	$[\mathbb{H}]$ is the $L_xL_yL_z \times L_xL_yL_z$ matrix 
	that has complex-valued random numbers in its diagonal elements and Hermitian hoppings with the unit modulus 
	in its off-diagonal elements. The cubic lattice is decomposed into A sublattice and B sublattice, where 
	the off-diagonal elements appear only between these two sublattices. Let the $L_xL_yL_z$-component 
	eigenvector $\Psi$ be transformed by a diagonal matrix $[\mathbb{B}]$ that has $-{\rm i}$ for B sublattice 
	and $+1$ for A sublattice;
	\begin{align}
		\Psi  = [\mathbb{B}] \Phi.    \label{relation1}
	\end{align}
	Since $E=0$, let another diagonal matrix $[\mathbb{A}]$ apply from the left of Eq.~(\ref{E=0}). 
	$[\mathbb{A}]$ has $+{\rm i}$ for A sublattice and $+1$ for B sublattice;
	\begin{align}
		[\mathbb{A}\mathbb{H}\mathbb{B}] \Phi = 0 \label{E=0d}
	\end{align}
	Now that the hopping terms appear only between A and B sublattices, 
	the off-diagonal terms in $\mathbb{H}^{\prime}\equiv \mathbb{A}\mathbb{H}\mathbb{B}$ are
	identical to those in $\mathbb{H}$. Meanwhile the real and imaginary parts of the diagonal elements in 
	$\mathbb{H}^{\prime}$ are exchanged with each other, compared to those in $\mathbb{H}$;
	\begin{eqnarray}
		\left\{\begin{array}{cc} 
			\mathbb{H}^{\prime}_{({\bm i},{\bm i})}  = {\rm i}  \mathbb{H}_{({\bm i},{\bm i})} & {\bm i} \in {\rm A}, \\
			\mathbb{H}^{\prime}_{({\bm i},{\bm i})} = - {\rm i} \mathbb{H}_{({\bm i},{\bm i})} & {\bm i} \in {\rm B}. \\
		\end{array}\right. \label{exchange}
	\end{eqnarray}
	$w_{\bm i}^r$ and $w_{\bm i}^i$ in $\mathbb{H}$ are uniformly 
	distributed in a range of $[-W_r/2,W_r/2]$ and $[-W_i/2,W_i/2]$, respectively. 
    Besides, on-site disorder potentials have no correlation between different 
    lattice sites. Thus, an ensemble of different disorder realization for 
    $\mathbb{H}$ with $(W_r,W_i)=(x,y)$ is the same as an ensemble of 
    different disorder realization for $\mathbb{A}\mathbb{H}\mathbb{B}$ 
    with $(W_r,W_i)=(y,x)$. Now that the zero-energy eigenfunction of 
    $\mathbb{H}$ and that of $\mathbb{A}\mathbb{H}\mathbb{B}$ are related to 
    each other by Eq.~(\ref{relation1}) and the transformation $\mathbb{B}$ 
    does not change localization lengths of these two eigenfunctions, 
    the phase diagram at $E=0$ must be symmetric with respect to an change between $W_i$ and 
    $W_r$. 
	
	One could also repeat the same argument 
	in the framework of the transfer matrix method. Thereby, one can introduce a
	product of another transfer matrix 
	\begin{align}
		M^{\prime}_{i_z} &\equiv \mathbb{A}_{i_z}  M_{i_z} \mathbb{B}_{i_z} 
		\equiv  \left(\begin{array}{cc} 
			\mathbb{a}_{i_z} & 0 \\
			0 & \mathbb{a}_{i_z-1} \\
		\end{array}\right) M_{i_z} \left(\begin{array}{cc} 
			\mathbb{b}_{i_z} & 0 \\
			0 & \mathbb{b}_{i_z-1} \\
		\end{array}\right),  \nonumber \\
		M^{\prime} & \equiv \prod^{L_z}_{i_z=1} M^{\prime}_{i_z} = \mathbb{A}_{L_z} M \mathbb{B}_{1}.  
		\label{transfer-1}
	\end{align} 
	Here $\mathbb{a}_{i_z}$ ($\mathbb{b}_{i_z}$)  is a $L_xL_y \times L_xL_y$ diagonal matrix 
	whose $(i_x,i_y)$ diagonal element takes $+{\rm i}$ ($-{\rm i}$) for $(i_x,i_y,i_z) \in$ A (B) sublattice and 
	takes $+1$ for $(i_x,i_y,i_z) \in$ B (A) sublattice. 
	Here we used $\mathbb{B}_{i_z} \mathbb{A}_{i_z-1}={\mathbb 1}$. 
	Eq.~(\ref{transfer-1}) dictates that the Lyapunov exponents of $M$ are identical 
	to those of $M^{\prime}$. 
	Note the hoppings terms in $M^{\prime}_{i_z}$ are the same as 
	those in ${M}_{i_z}$, and the real and imaginary parts of the on-site NH disorder potentials in $M^{\prime}_{i_z}$ 
	are exchanged with each other, compared to those in $M_{i_z}$; Eq.~(\ref{exchange}). 
	We therefore obtain the same results at $E=0$ for $W_r=0$ and for $W_i=0$.
	
	When $E\ne 0$, the phase diagram of the three NH models becomes 
	asymmetric with respect to the exchange of $W_r$ and $W_i$. 
	Besides, the critical point along an axis of $W_r=0$ at $E\ne 0$ belongs {\it not} to the 
	3D Hermitian orthogonal/unitary class, but to the new universality classes 
	AI$^{\dagger}$/A of the NH systems. In the following three subsections, 
	we first clarify critical behaviors of the Anderson transition in the NH Anderson model.

	\subsection{critical behaviors of the Anderson transition in the complex energy plane (NH Anderson model)}
	
	\begin{figure*}[tb]
		\centering
		\subfigure[ $E=0$ ]{
			\begin{minipage}[t]{0.5\linewidth}
				\centering
				\includegraphics[width=1\linewidth]{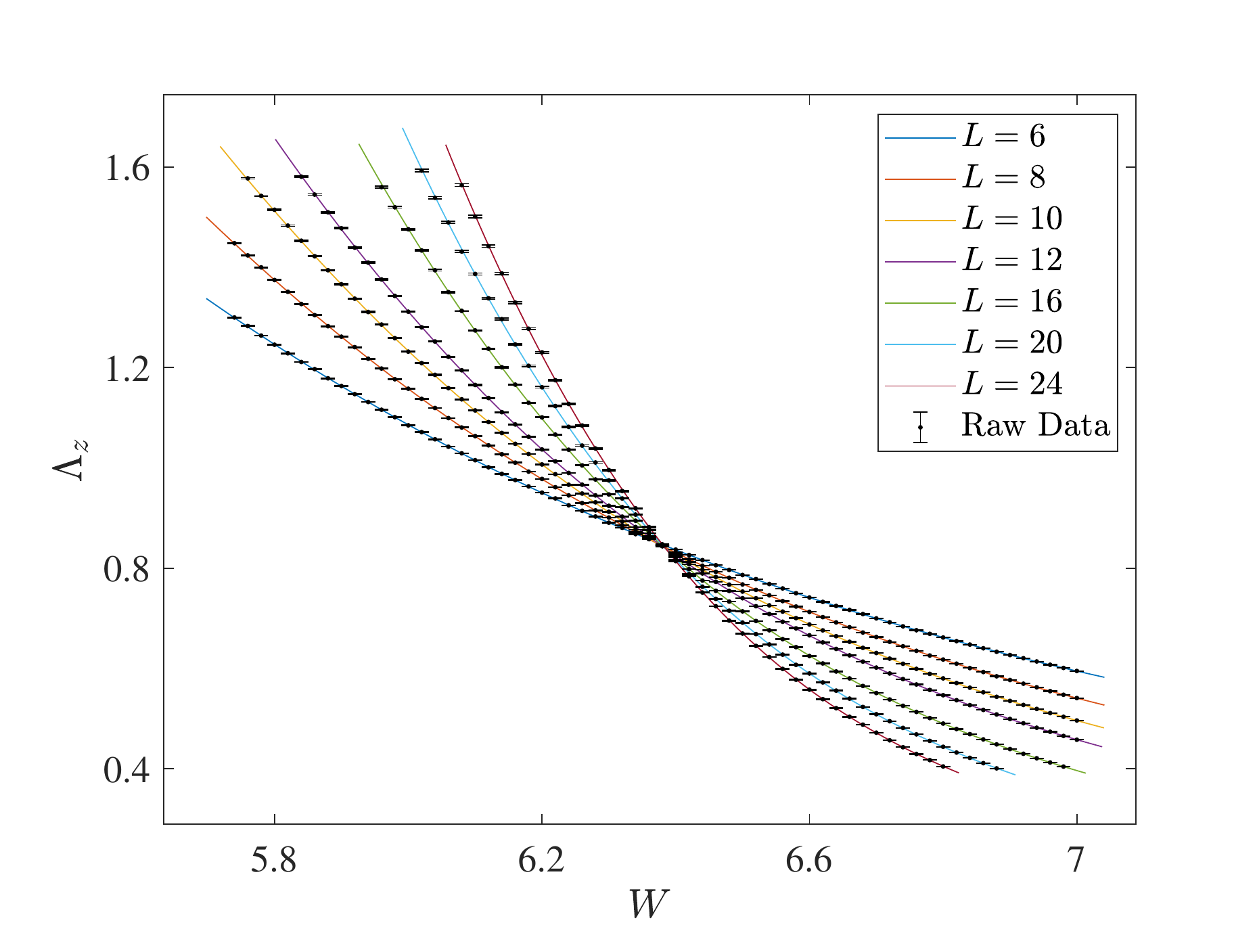}
			\end{minipage}%
		}%
		\subfigure[ $E={\rm i}$ ]{
			\begin{minipage}[t]{0.5\linewidth}
				\centering
				\includegraphics[width=1\linewidth]{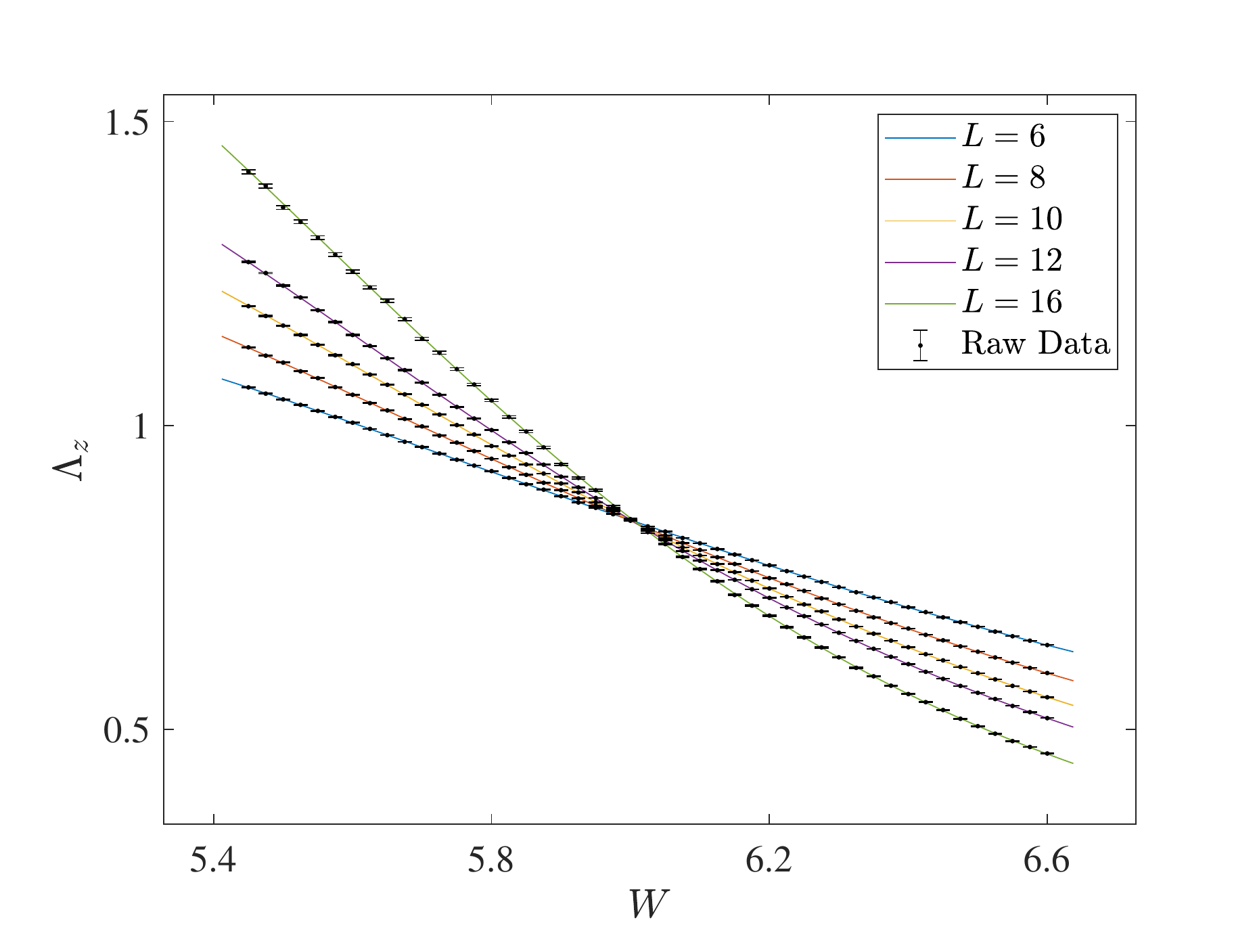}
			\end{minipage}%
		}%
		\caption{ Polynomial fitting of $\Lambda_z$ at (a) $ E=0$ and (b) $E={\rm i}$ with $W_r=W_i=W$ for Anderson model. 
			The black points with error bar are the raw data of $\Lambda_z$ and lines with different colors  
			are the polynomial fitting results with expansion order ($m_1$, $n_1$,  $m_2$,  $n_2$)=(3,3,0,1).}
		\label{E0E1i}
	\end{figure*}
	
	\begin{figure}
		\centering
		\includegraphics[width=1\linewidth]{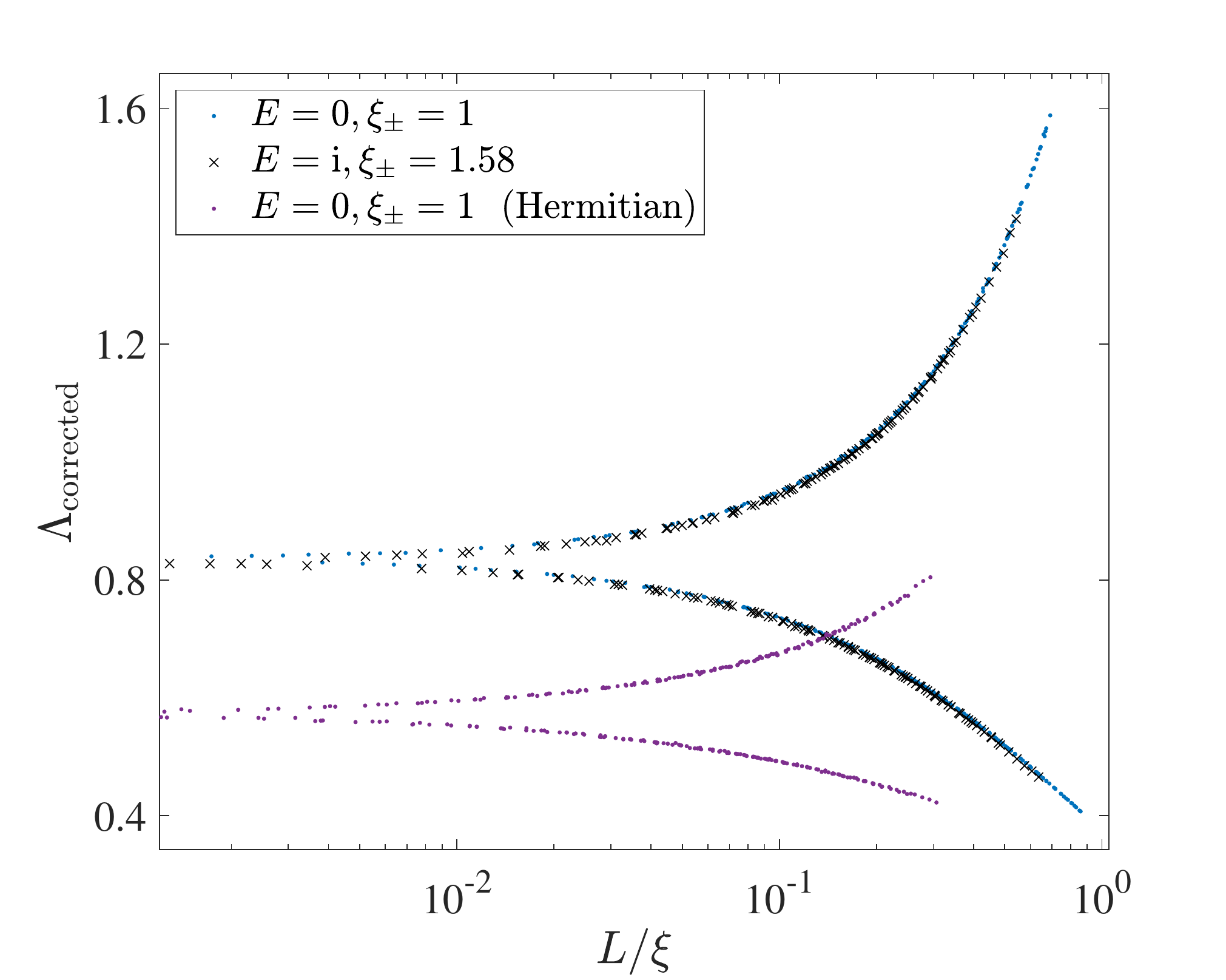}
		\caption{
			The data in Fig. \ref{E0E1i} after subtraction of correction [see Eq. (\ref{Lambda_corrected})] collapse into one curve according to scaling function Eq. (\ref{Lambda_scaling}). Here we set $\xi_{\pm}=1$ for $E=0$ and $\xi_{\pm}=1.58$ for $E={\rm i}$. The data of the Hermitian case at $E=0$ are also shown for the comparison.}
		\label{fig_scaling}
	\end{figure}
	
	In this subsection, we first compare critical properties of the Anderson transition 
	at different energies in the complex plane. Eigenvalues of the NH systems are generally 
	complex numbers and the eigenvalues of the NH Anderson model are distributed 
	around $E=0$ in the Euler plane. In the Hermitian Anderson model, 
	the Anderson transition for different energies appears at different critical disorder strength 
	with the same critical exponent. A natural question arises in the NH Anderson model, asking whether 
	the critical behaviors of the Anderson transition at different $E$ in the complex plane are the same 
	or not. To answer this question, we calculate the localization length at $E=0$ and $E={\rm i}$, while changing 
	the disorder strength $W$ along the symmetric line in the $W_i$-$W_r$ plane; $W_r=W_i=W$. 
	
	The localization length is calculated at $E=0$ with $L_z=10^7$ for $L=6,8,10,12,16$, 
	$L_z=6\times 10^6$ for $L=20,24$; Fig. \ref{E0E1i}(a), and at $E={\rm i}$ with 
	$L_z=10^7$ for $L=6,8,10,12$, $L_z=2\times 10^6$ for $L=16$; Fig. \ref{E0E1i}(b). 
	Since there are no eigenstates at $E={\rm i}$ at $W=0$, the system at $E={\rm i}$ first 
	undergoes a localization-delocalization transition at smaller critical disorder strength 
	and then undergoes another delocalization-localization transition at larger critical disorder strength 
	($W_c\approx 6.02$). The two-steps transition at $E={\rm i}$ is similar to a reentrant phenomenon 
	observed in Hermitian Anderson model near band edges\cite{Bulka87,Kramer93}.
	The density of states at $E={\rm i}$ is always finite around these two transition 
	points. It is also finite at the Anderson transition at $E=0$ ($W_c\approx 6.39$). Let us compare 
	the critical behaviors of $E=0$ at $W_c\approx 6.39$ with those of $E={\rm i}$ at $W_c\approx 6.02$.  
	
	The polynomial fitting method is used for the purpose of extracting critical quantities; Table~\ref{table}. 
	The critical disorder strength at $E=0$ is larger than those at $E={
		\rm i}$; 
	delocalized states at the center of the complex plane is more robust than the states otherwise. 
	The critical exponent $\nu\approx 1.19$, 
	critical normalized localization length $\Lambda_c\approx 0.83$, and the least irrelevant scaling variable 
	$y\approx 2.5$ are all consistent with each other between these two transition points. 
	The coincidence suggests that the Anderson transition at different energies in the complex energy 
	plane shares the same critical behaviors. $\nu$ and $\Lambda_c$ at these two points
	are distinct from those values of the 3D orthogonal class in the Hermitian case; 
	$\nu\approx 1.57$ and $\Lambda_c\approx 0.58$ \cite{Slevin14}. 
	
	In order to determine a universal scaling function form of $\Lambda$ 
	at the NH Anderson transition point, we subtract $\Lambda$ by the finite-size correction 
	due to the irrelevant scaling variable,
	\begin{equation}\label{Lambda_corrected}
		\Lambda_{\rm corrected} \equiv \Lambda- \big[F(\phi_1,\phi_2)-F(\phi_1,0)\big].
	\end{equation} 
	According to Sec. \ref{subsec:polynomiaFitting}, the corrected $\Lambda$ must be given by a 
	universal single-parameter scaling function near the transition point, 
	\begin{equation}\label{Lambda_scaling}
		\Lambda_{\rm corrected}=f_{\pm}\Big(\frac{L}{\xi}\Big),
	\end{equation}
	where $\xi\equiv\xi_{\pm} |u_1(\omega)|^{-\nu}$.
	Here $\xi_{\pm}$ depends on non-universal quantities in 
	the polynomial fitting analysis, such as $a_{1,1}$, 
	and they generally take different values at different single-particle 
	energies and system parameters. 
	However, $\xi_+/\xi_-$ would take a universal value \cite{Privman91}.
	Fig.~\ref{fig_scaling} shows that with a proper choice of $\xi_{\pm}$, 
	all the data points for $E=0$ and those for $E={\rm i}$ near the respective 
	transition points fall into a single curve, demonstrating the validity of the universal 
	single-parameter scaling function form for the corrected $\Lambda$. 
	The single-parameter scaling function around the NH Anderson 
	transition point is clearly distinct from that around the Hermitian Anderson transition point 
	(Fig.~\ref{fig_scaling}). This unambiguously shows that the Anderson transition in the NH 
	system belongs to a new universality class \cite{Luo21}.

	\subsection{critical behaviors of the Anderson transition in the phase diagram at $E=0$ (NH Anderson model)}\label{PD_E0}
	
	\begin{figure*}[tb]
		\centering
		\subfigure[$W_i=5$, (3,3,0,1) ]{
			\begin{minipage}[t]{0.5\linewidth}
				\centering
				\includegraphics[width=1\linewidth]{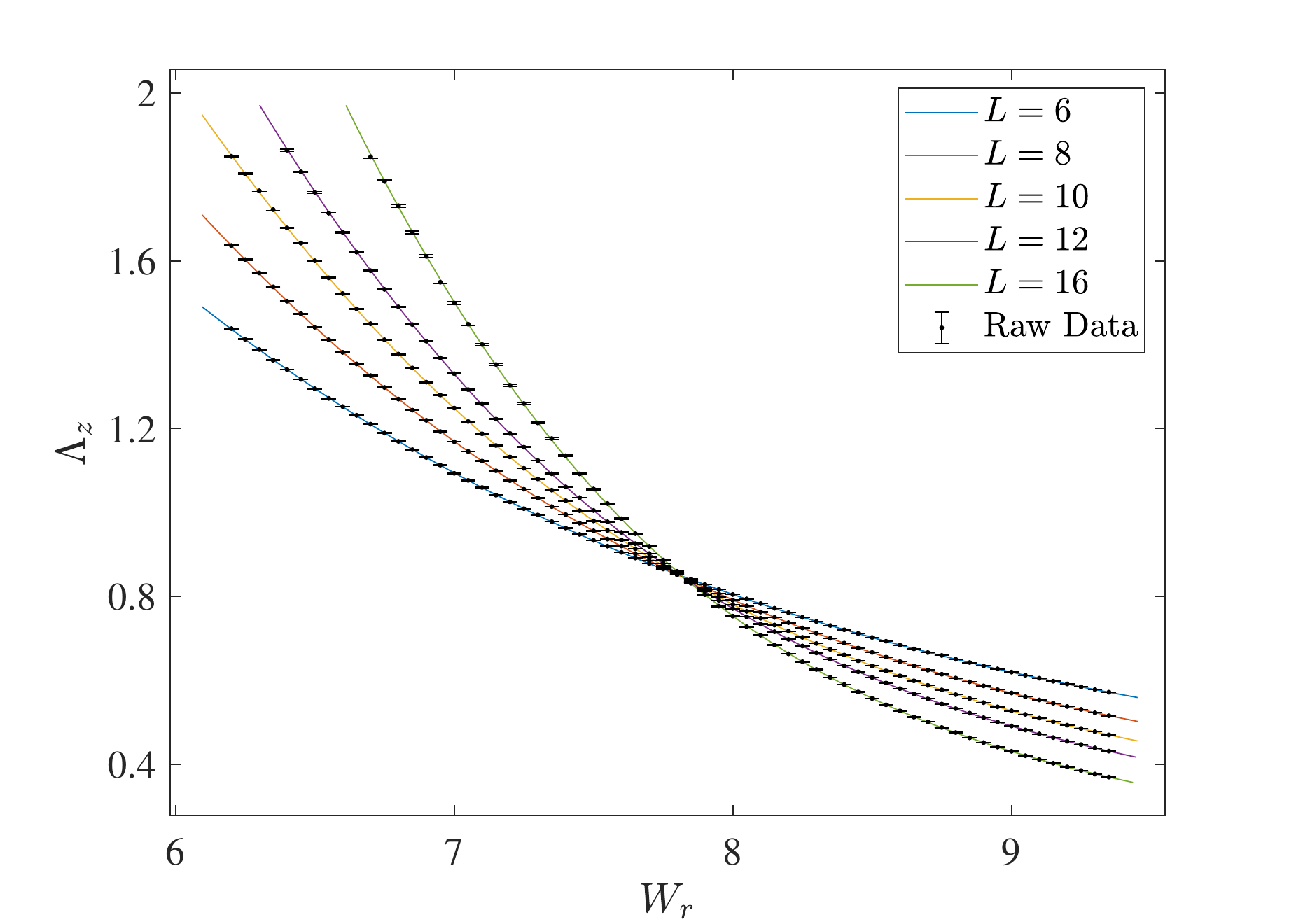}
			\end{minipage}%
		}%
		\subfigure[$W_r=5$, (3,3,0,1) ]{
			\begin{minipage}[t]{0.5\linewidth}
				\centering
				\includegraphics[width=1\linewidth]{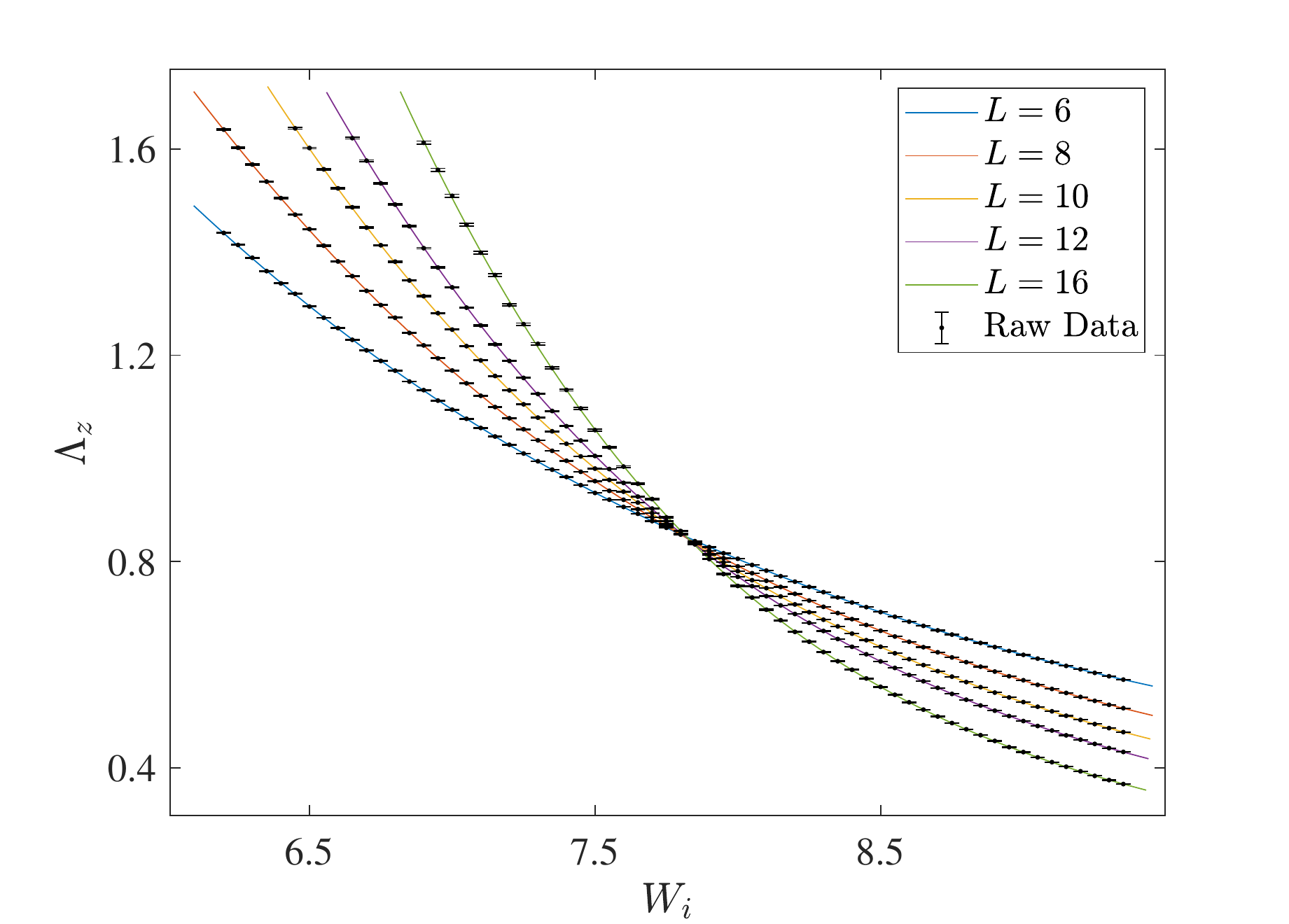}
			\end{minipage}%
		}%
		
		\subfigure[$W_i=0.1$, (3,3,1,1) ]{
			\begin{minipage}[t]{0.5\linewidth}
				\centering
				\includegraphics[width=1\linewidth]{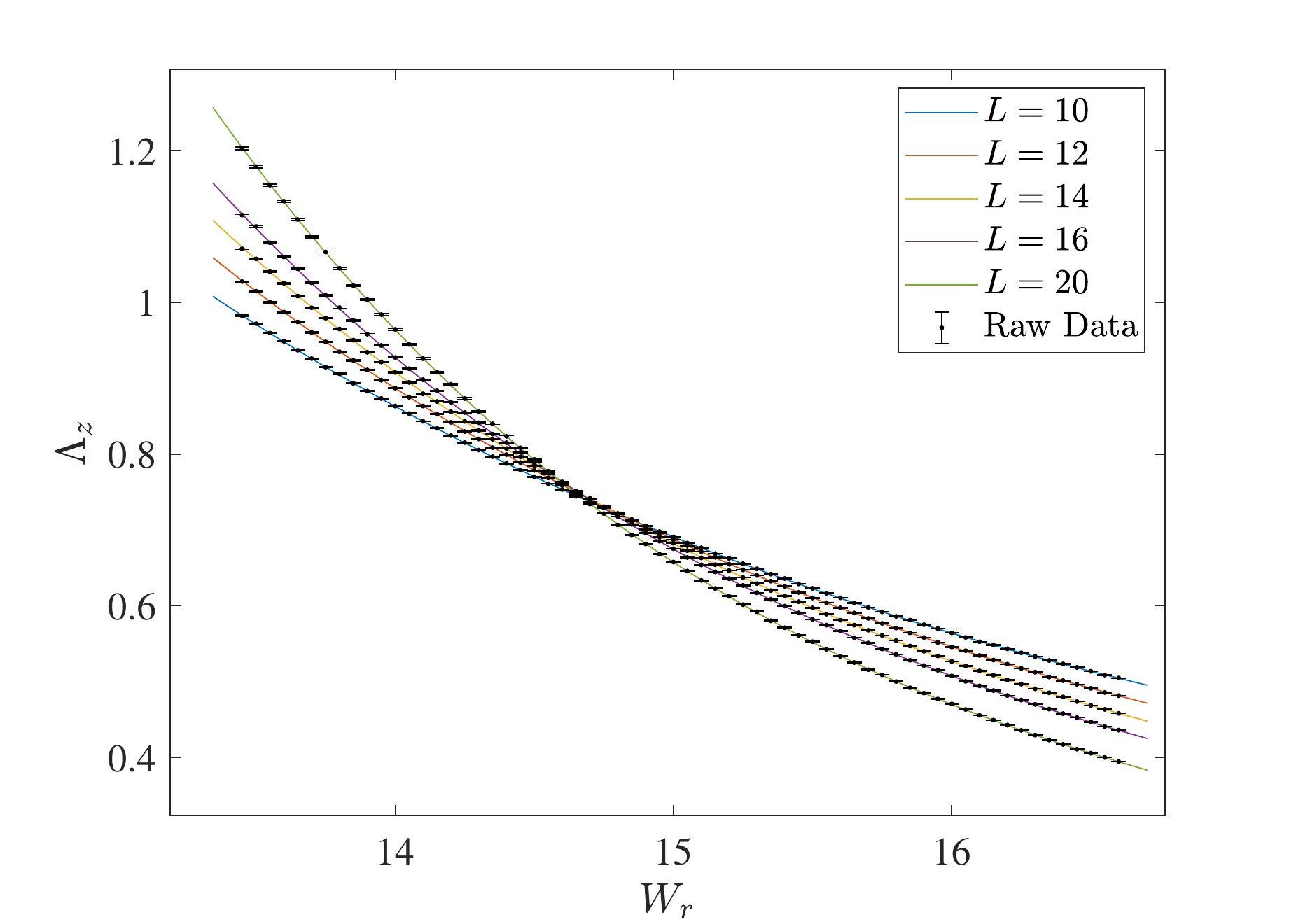}
			\end{minipage}%
		}%
		\subfigure[ $W_r=0$, (2,3,0,1)]{
			\begin{minipage}[t]{0.5\linewidth}
				\centering
				\includegraphics[width=1\linewidth]{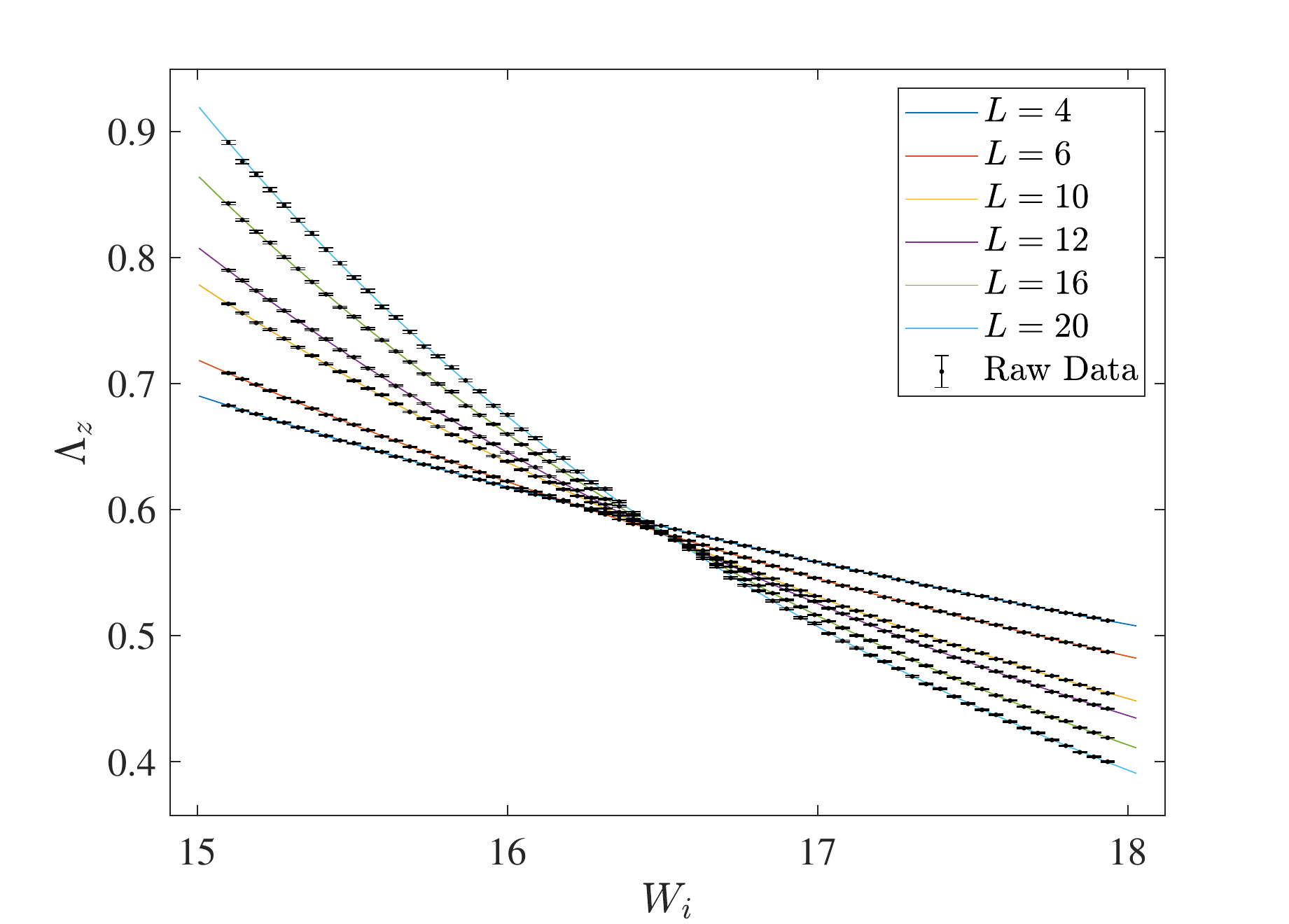}
			\end{minipage}%
		}%
		\caption{ Polynomial fitting for $\Lambda_z$ at $E=0$ with fixed $W_r$ or $W_i$ for the NH Anderson model. The black points with error bar are the raw data of $\Lambda_z$ and lines with different colors 
			are the polynomial fitting results with expansion order ($m_1$, $n_1$,  $m_2$,  $n_2$).
			The expansion order ($m_1$, $n_1$,  $m_2$,  $n_2$) are shown in the brackets for each figure.}
		\label{Wi_WrFixed}
	\end{figure*}
	In this subsection, we compare critical properties of the Anderson transition 
	at different system parameters of the same energy $E=0$. 
	The localization length is calculated as a function of $W_r$ for fixed $W_i=5$; Fig. \ref{Wi_WrFixed}(a)  
	or as a function of $W_i$ for fixed $W_r=5$; Fig. \ref{Wi_WrFixed}(b), where $L_z=10^7$ for 
	$L=6,8,10,12$ and $L_z=3\times10^6$ for $L=16$. The polynomial fitting results are summarized in the 
	Table \ref{table}. In the Table, the critical exponent $\nu\approx1.19$, the critical length 
	$\Lambda_c\approx0.83$, the least irrelevant scaling dimension and the critical disorder 
	are consistent with each other for the fixed $W_i=5$ case and for the 
	fixed $W_r=5$ case. Besides, the critical exponent and the critical length for these two 
	transition points are consistent with those transition points at $E=0$ and $E={\rm i}$ along 
	$W_r=W_i=W$. These comprehensive analyses conclude that the critical exponent 
	for the Anderson transition in the NH Anderson model is
	\begin{equation}\label{CE}
		\nu=1.192[1.185, 1.198].
	\end{equation}
	
	The localization length at $E=0$ is also calculated as a function of $W_i$ for the fixed $W_r=0$ with $L_z=10^7$ for $L=4,6,10,12,16$, and $L_z=5\times10^6$ for $L=20$; Fig.~\ref{Wi_WrFixed}(d). Polynomial fitting results are shown in 
	Table~\ref{table}. As expected from the symmetry argument above, the critical 
	exponent $\nu\approx1.57$, the critical length $\Lambda_c\approx0.58$ and and the critical 
	disorder $W_c\approx16.5$ coincide with those values 
	in the Hermitian limit ($W_i=0$).

	\subsection{critical behaviors of the Anderson transition at $W_r=0$ and $E \ne 0$ (NH Anderson model)}\label{PD_E_ne_zero}
\begin{figure}[tb]
		\centering
		\includegraphics[width=1.0\linewidth]{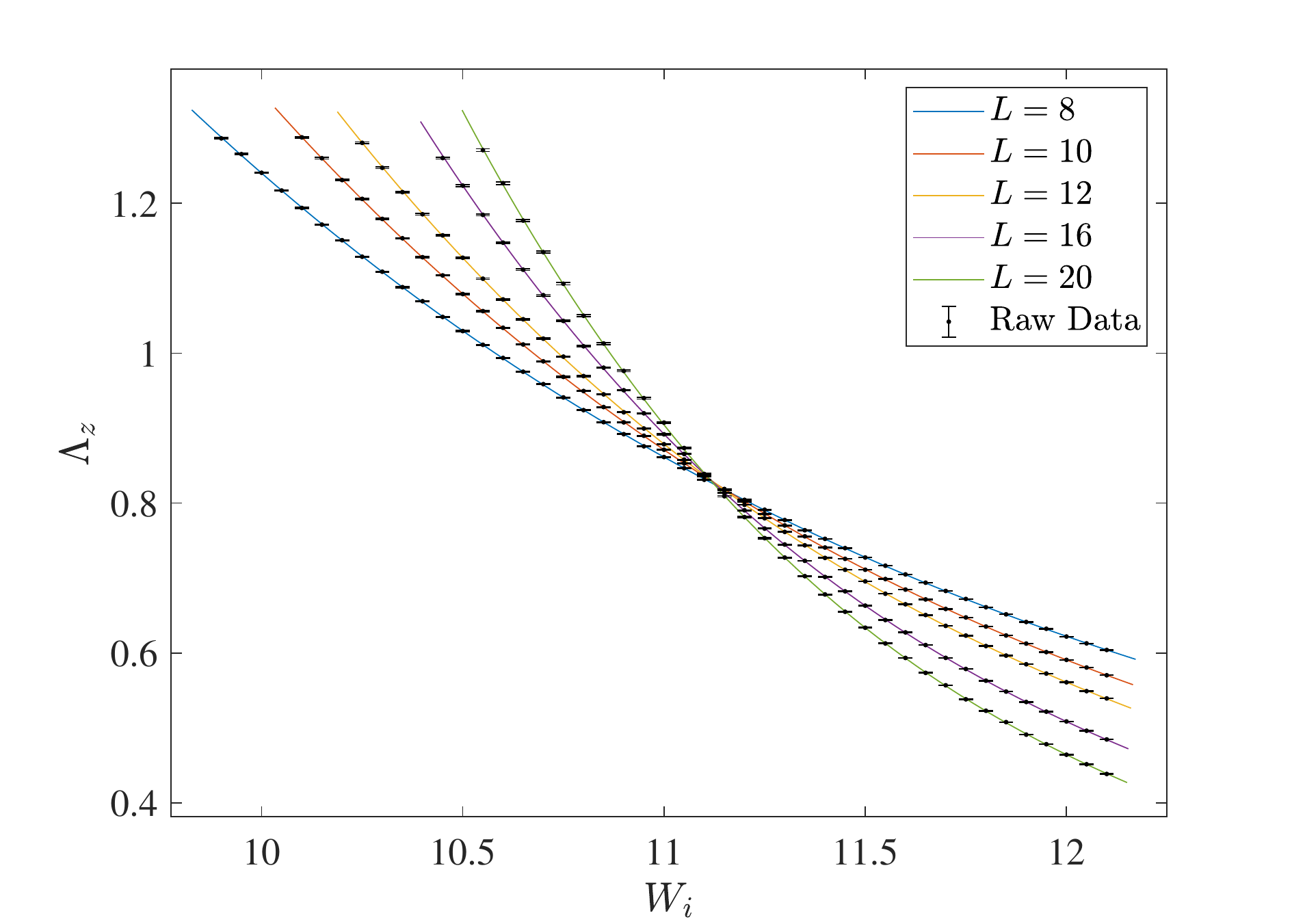}
		\caption{Polynomial fitting of $\Lambda_z$ at $E=2$ and $W_r=0$ for the NH Anderson model.
			The black points with error bar are the raw data of $\Lambda_z$
			and lines with different colors are the polynomial fitting results with expansion order 
			($m_1, n_1, m_2, n_2$)=(2,3,0,1).}
		\label{figE=2}
	\end{figure}

	The symmetric nature of the phase diagram and critical properties with respect to the exchange of 
	$W_r$ and $W_i$ is absent at $E\ne 0$. Thereby, the critical point along the axis of $W_r=0$ 
	must belong to the same universality class as the NH Anderson model. 
	To confirm this, we calculate the localization length at $W_r=0$ and $E=2$ with 
	$L_z=10^7$ for $L=8,10,12,16$ and with $L_z=6\times 10^6$ for $L=20$; Fig.~\ref{figE=2}.  
	Polynomial fitting results are summarized in the Table \ref{table}. The critical disorder strength 
	($W_c\approx11.1$) is clearly different from the value in the Hermitian limit ($W_c\approx16.5$), 
	demonstrating the asymmetry of the phase diagram at $E=2$. Both the critical exponent $\nu\approx1.20$ 
	and the critical length $\Lambda_c\approx0.83$ are consistent with those values from 
	$E=0$ and $E={\rm i}$ along $W_r=W_i=W$ and from $E=0$ along $W_{r(i)}=5$.  
	The result reinforces the conclusion of Eq.~(\ref{CE}).  
	
	\subsection{RG flows at $E=0$ and at $E\ne 0$ and NH-H crossover phenomena (NH Anderson model)}
	In this subsection, we postulate an RG flow diagram in the $W_r$-$W_i$ plane at 
	$E=0$ and at $E\ne 0$, based on the critical properties summarized in Sec. \ref{PD_E0} and 
	Sec. \ref{PD_E_ne_zero}; Fig. \ref{phasediagram}(b, c). 
	The RG flow diagram at $E=0$ has a saddle-point fixed point (FP3) with only one relevant scaling 
	variable and two unstable fixed points (FP1 and FP2) with two relevant scaling variables. 
	The FP3 determines the critical properties of delocalization--localization transition
	in the NH systems. When a simulated lattice model 
	goes across the phase transition line, any dimensionless physical quantity is given by the universal 
	function of the relevant scaling variable around FP3 and other irrelevant scaling variables. One of the 
	irrelevant scaling variables is shown explicitly in the RG flow diagram in the $W_r$-$W_i$ plane. 
	
	The FP1 (FP2) controls the criticality in the Hermitian limit. When the lattice model crosses the 
	transition line with either $W_r=0$ or $W_i=0$ at $E=0$, one of the two relevant scaling variables 
	around FP1 (FP2) can be always set to zero. This is because the renormalization must respect 
	the symmetry of the system and therefore it does not change a Hermitian system into a 
	NH system. As a result, the quantity can be given by the universal function of only the 
	other of the two relevant variables around FP1 (FP2); the Hermitian criticality.  In the RG flow 
	diagram at $E\ne 0$, FP1 disappears in the $W_r$-$W_i$ plane and all the critical properties 
	except for those along the axis of $W_i=0$ are controlled by FP3; Fig. \ref{phasediagram}(c). 
	
	When the model undergoes the transition near FP2, the critical properties might  
	exhibit a crossover phenomenon from the Hermitian case to the NH case. To 
	study this crossover phenomenon, we calculate the localization length at $E=0$ for fixed 
	$W_i=0.1$ with $L_z=10^7$ for $L=10,12,14,16$ and $L_z=6\times 10^6$ for $L=20$; 
	Fig. \ref{Wi_WrFixed}(c). The fitting results are summarized in Table~\ref{table}. According 
	to the fitting results, the scaling dimension of the least irrelevant scaling variable is 
	quite small, indicating the importance of the irrelevant scaling variables. In fact, the large correction 
	by the irrelevant scaling variables results in a bad intersection of $\Lambda_z$ in the data. 
	The critical exponent in the fitting results is consistent neither with that for the Hermitian case 
	nor with that for the NH case. To capture the correct critical properties in this crossover regime, 
	we believe it necessary either to include the non-linear $\phi_2$-dependence of $F$ ($n_2 > 1$) 
	together with the non-linear $w$-dependence of $u_2(w)$ ($m_2>1$) or to fit only the 
	larger system size data.
	
	\subsection{conductance (NH Anderson model)}
	\label{sec:conductance}
	
	As in the Hermitian case \cite{Slevin01,Slevin03}, the conductance is a convenient physical quantity 
	that can characterize the critical property of the Anderson transition in the 
	NH systems. As shown in the previous section, the conductance in the NH systems 
	is not bounded by the total number of the transmission channels in the two-terminal 
	geometry. The transmission amplitude can be arbitrarily amplified by the on-site NH 
	disorder potentials in the NH systems. 
	
\begin{figure*}[tb]
		\centering
		\subfigure[$W=1$, metal phase]{
			\begin{minipage}[t]{0.33\linewidth}
				\centering
				\includegraphics[width=1\linewidth]{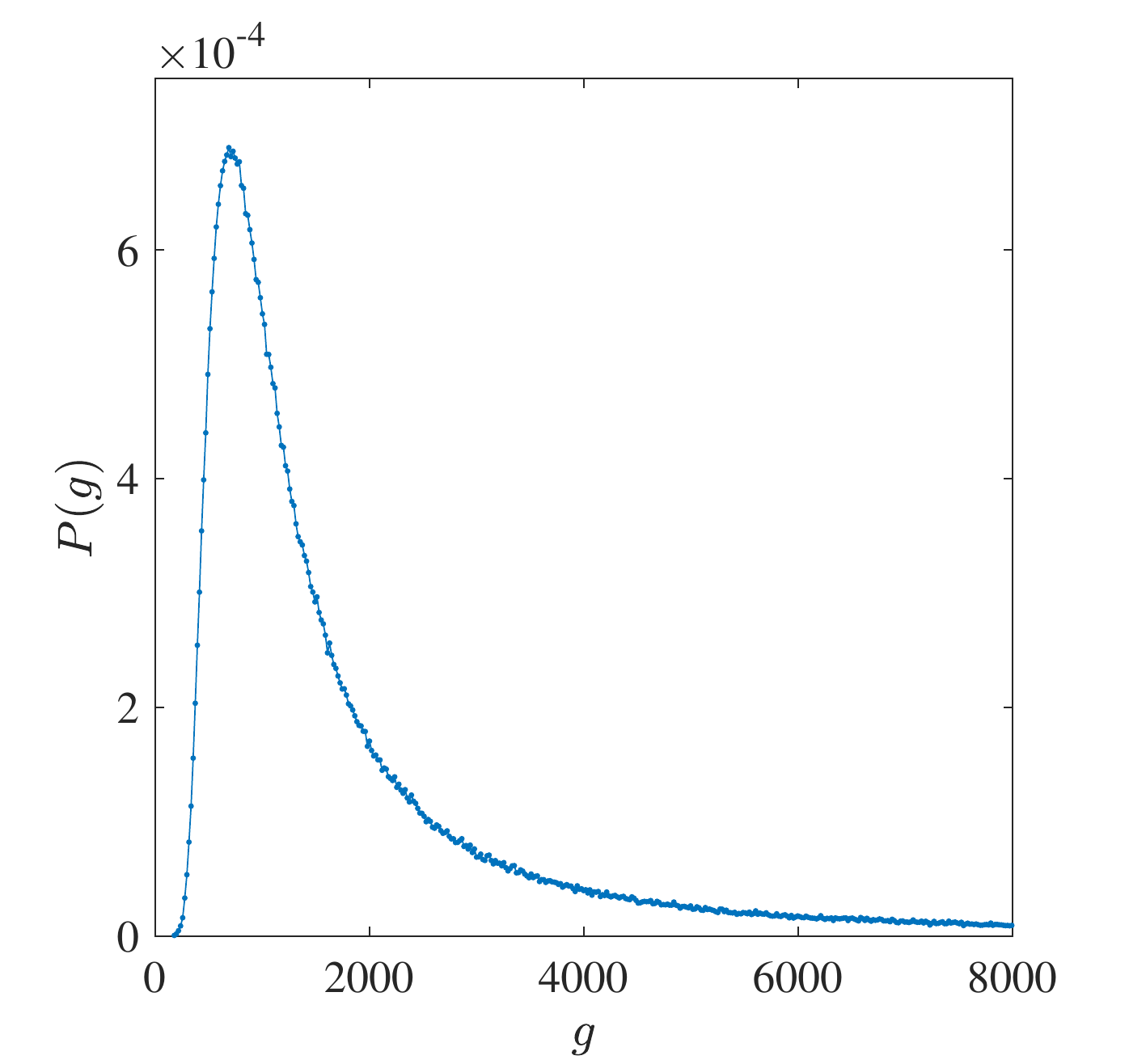}
			\end{minipage}%
		}%
		\subfigure[$W=6.386$, near critical point]{
			\begin{minipage}[t]{0.33\linewidth}
				\centering
				\includegraphics[width=1\linewidth]{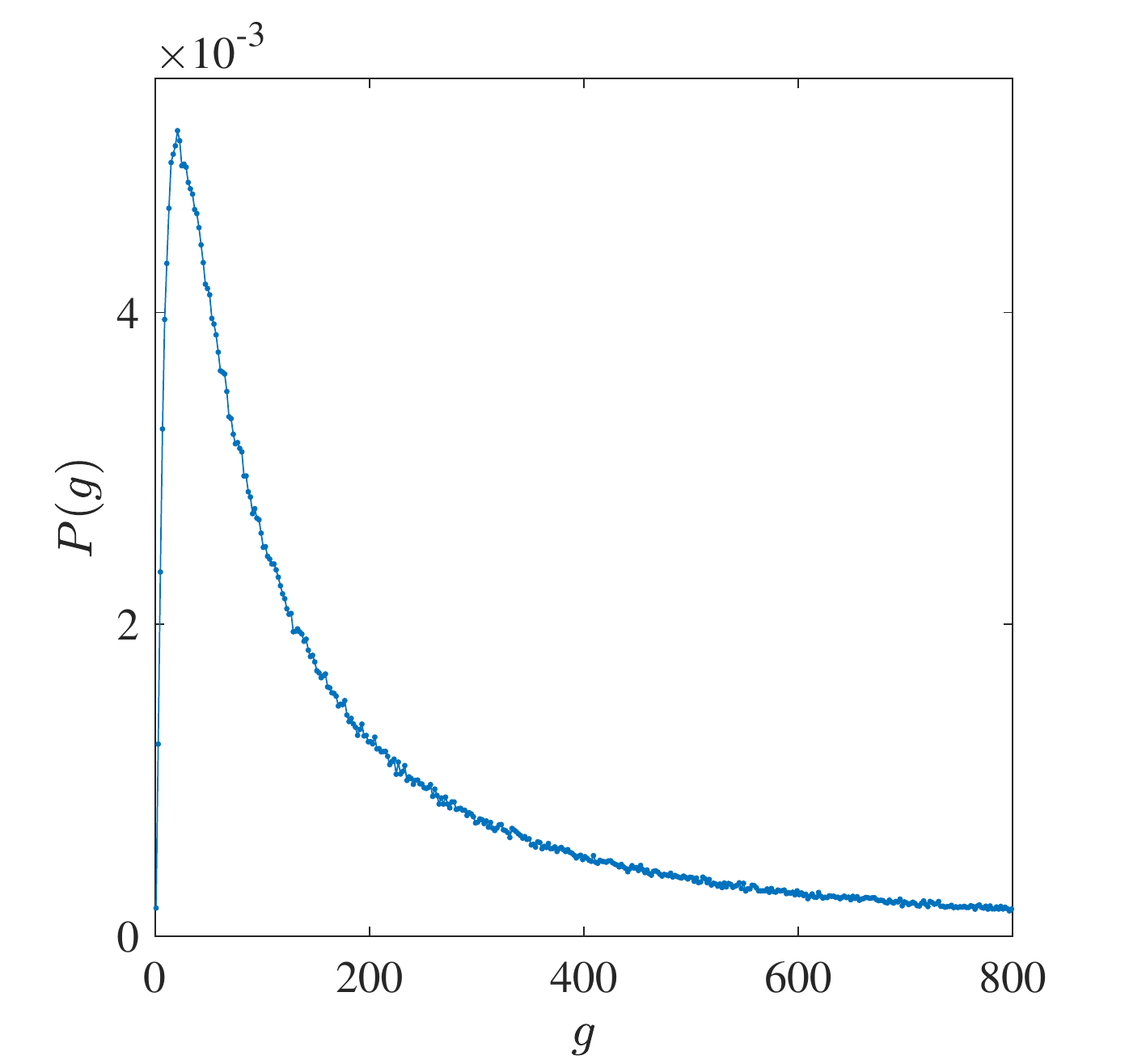}
			\end{minipage}%
		}%
		\subfigure[$W=14$, insulator phase]{
			\begin{minipage}[t]{0.33\linewidth}
				\centering
				\includegraphics[width=1\linewidth]{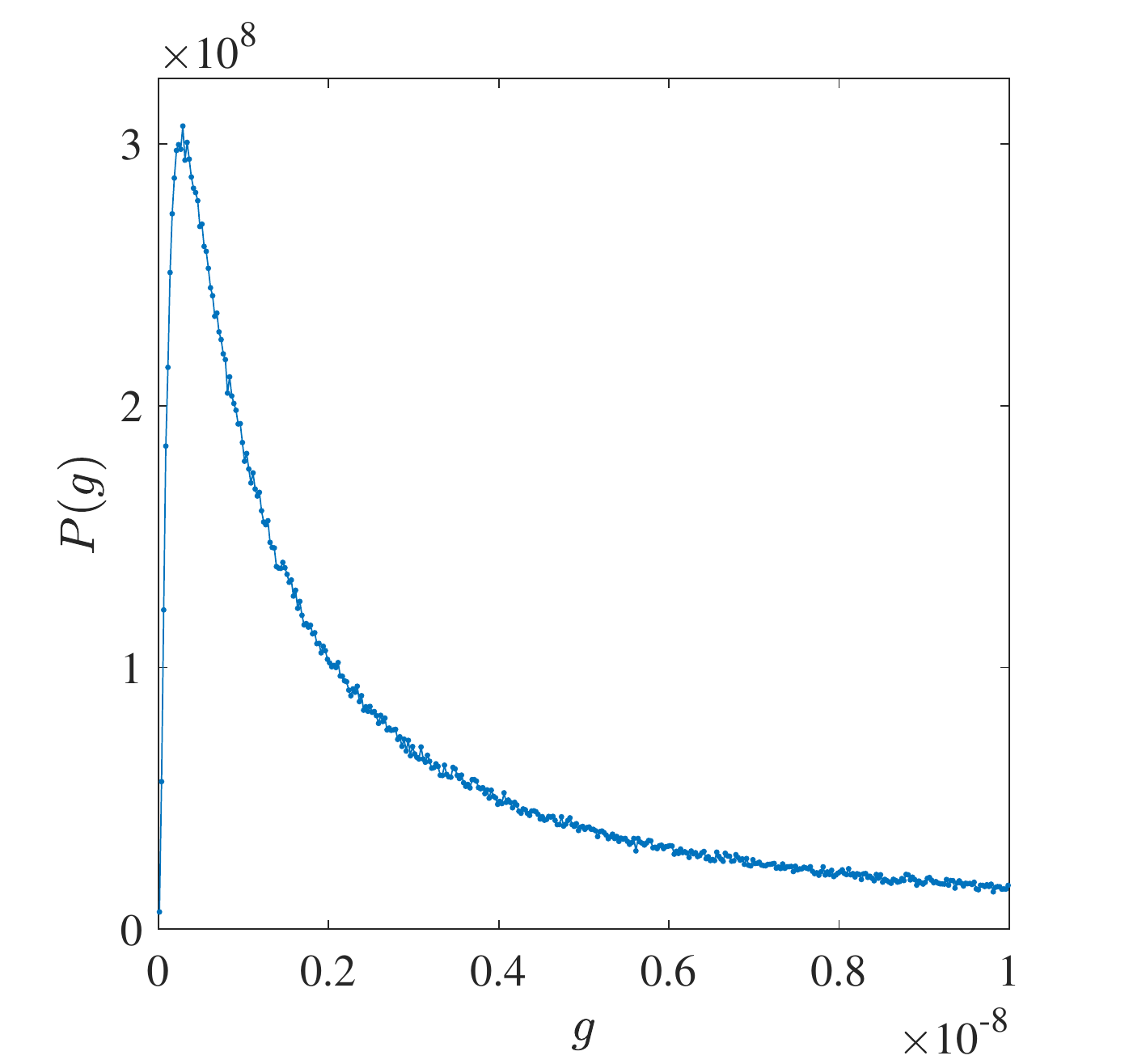}
			\end{minipage}%
		}%
		\caption{Conductance distribution $P(g)$ of the Anderson model with non-Hermitian disorder. 
			$10^6$ samples are calculated for $L=12$ at $E=0$ and $W_r=W_i=W$. (a) $W=1$: metallic 
			phase, (b) $W=6.386$: critical point, and (c) $W=14$: localized phase. The distribution has a long 
			tail in a region of very large conductance value. The horizontal axis is an arbitrary unit. Only 
			$93\%, 83\%, 68\%$ of the whole data points are shown in (a,b,c), respectively.}
		\label{g_dist}
	\end{figure*}

	Fig. \ref{g_dist} shows distributions of 
	the conductance in the NH Anderson models, where the conductance is 
	calculated in terms of Eqs.~(\ref{two-terminal-conductance}), 
	(\ref{whatisT}), and (\ref{T-def}) with the cubic geometry 
	$L_x=L_y=L_z=L$ for $10^6$ samples of different disorder realizations. 
	We have changed the frequency of the $QR$ decomposition in the 
	transfer matrix calculation, and checked the results in Fig.~\ref{g_dist} are 
	robust against the change. The three figures in Fig.~\ref{g_dist} show 
	that the distribution of the conductance is not Gaussian irrespective of 
	whether the system is in the delocalized phase or in the localized phase or at the 
	critical point. The conductance distribution always contains small fractions 
	of huge conductance values. The very large  
	conductance values come from `rare-event' states, in which the 
	transmission is strongly 
	amplified by the NH disorders. To carry out the FSS analysis for such 
	conductance data, we take a geometric mean of the conductance; $\langle \ln g \rangle$. 
	The plot of the geometric average of the conductance and 
	its fit to polynomial functions are shown in Appendix.
	
	\section{non-Hermitian class A models}
	\label{sec:NHU1}
	In this section, we give an accurate characterization of the critical properties of the Anderson 
	transition in the two NH class A models; NH U(1) and Peierls models. From 
	the symmetry argument above, the $E=0$ phase diagrams of these two models must also be 
	symmetric with respect to the exchange of $W_r$ and $W_i$. At $E=0$, both the axis 
	of $W_r=0$ and the axis of $W_i=0$ belong to the Hermitian class A. We expect that an 
	effect of the irrelevant scaling variable can be minimized along the symmetry 
	line of $W_r=W_i$. Based on this anticipation, we focus our study on the Anderson transition 
	in the two NH class A models along the axis of $W_r=W_i=W$, setting $E=0$.
	
	\subsection{NH U(1) model at $E=0$}
\begin{figure}[tb]
		\centering
		\includegraphics[width=1\linewidth]{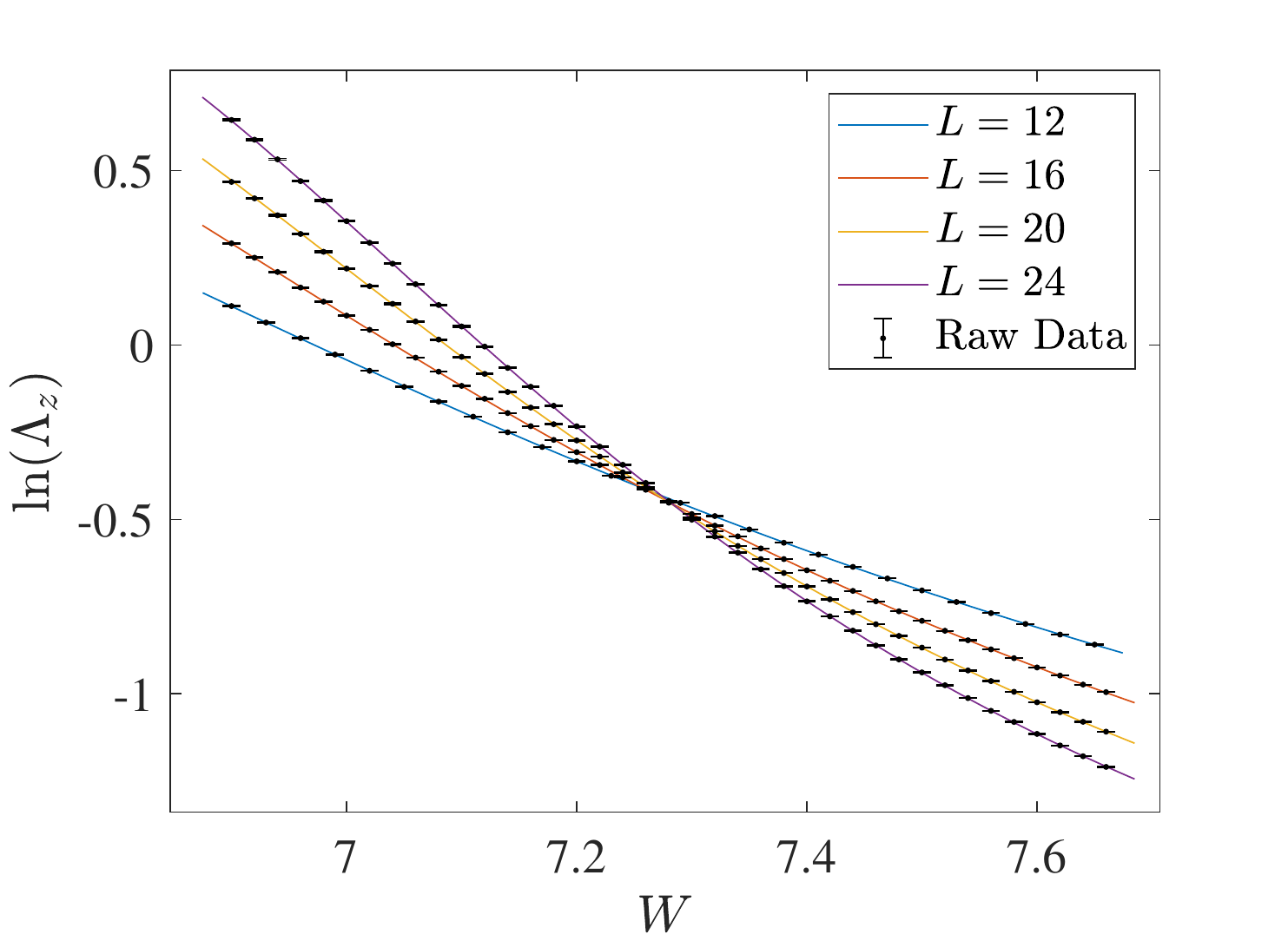}
		\caption{ Polynomial fitting of $\ln(\Lambda_z)$ at $E=0$ with $W_r=W_i=W$ for the NH U(1) model. The black points with error bar are the raw data of $\ln(\Lambda_z)$ and lines with different colors are the polynomial fitting results with expansion order ($m_1$, $n_1$, $m_2$,  $n_2$)=(3,3,0,1).}
		\label{u1}
	\end{figure}

	In this subsection, we first clarify the critical property of the Anderson transition 
	in the NH U(1) model. The localization length has been calculated with 
	$L_z=10^7$ for $L=8,10,12,16$, and with $L_z=6\times 10^6$ for $L=20, 24$.  
	For the purpose of the transfer matrix calculations, it is convenient to perform a 
	gauge transformation of the original Hamiltonian so as to eliminate all the random 
	U(1) phase factors appearing in hopping elements in the $z$-direction. Obviously, the 
	transformation does not affect the values of the Lyapunov exponents \cite{Slevin16}. 
	The normalized localization length near its scale invariant point is shown in Fig.~\ref{u1}. 
	The polynomial fitting result is summarized in Table~\ref{table}. The critical exponent is 
	evaluated from the fitting as 
	\begin{equation}\label{CE2}
		\nu=1.003[0.985, 1.018],
	\end{equation}
	together with $\Lambda_c\approx 0.598$. These two universal quantities are clearly distinct 
	from $\nu\approx 1.44$ and $\Lambda_c\approx 0.55$ for the Hermitian U(1) 
	model \cite{Slevin16}. They are also different from the $\nu\approx 1.19$ and 
	$\Lambda_c\approx 0.83$ for the NH Anderson model (NH class AI$^\dagger$). 
	As was pointed out in Ref.~\onlinecite{Luo21}, $W_c\approx 7.3$ for the NH U(1) model 
	is significantly larger than $W_c$ for the NH Anderson model ($W_c\approx 6.4$), in spite of the 
	fact that the NH U(1) model has more random variables in its off-diagonal matrix elements
	than 
	the NH Anderson model. This means that the delocalization-localization transition in the NH 
	system is also of the quantum interference origin. 
	
	The two-terminal conductance $g$ is also 
	calculated in the NH U(1) model along the axis of $W_r=W_i=W$ at $E=0$. A plot of the 
	geometric average of the conductance and the details of the polynomial fitting are 
	given in the Appendix.
	
	\subsection{NH Peierls model at $E=0$}
\begin{figure}[tb]
		\centering
		\includegraphics[width=1.0\linewidth]{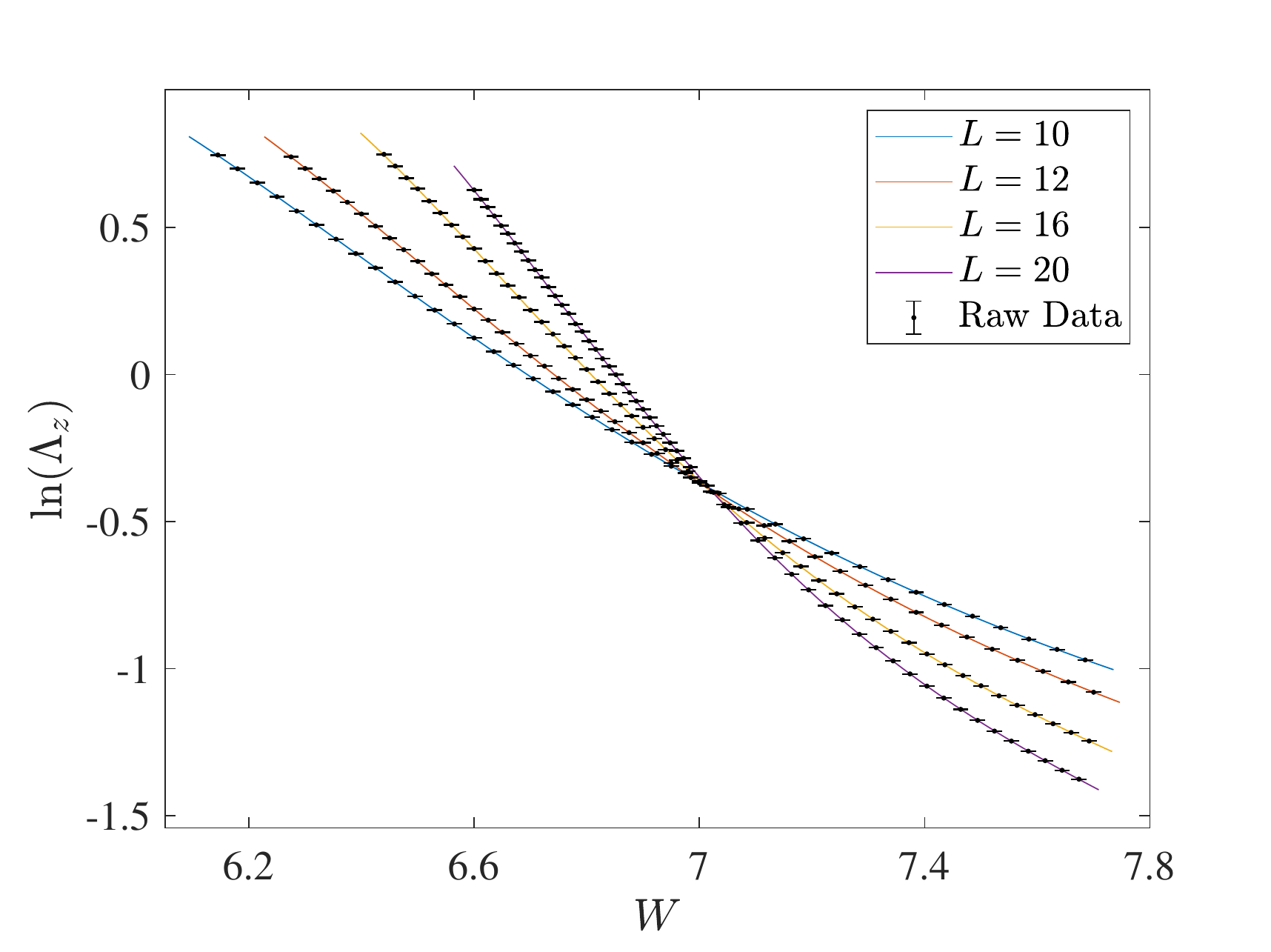}
		\caption{Polynomial fitting of $\ln (\Lambda_z)$ at $E=0$ and $W_r=W_i=W$ for the NH Peierls 
			model with the magnetic flux $\Phi=1/4$. 
			The black points with error bar are the raw data of $\ln(\Lambda_z)$ and
			lines with different colors are the polynomial fitting results 
			with expansion order ($m_1, n_1, m_2, n_2$)=(3,4,0,1).}
		\label{pie}
	\end{figure}

	In this subsection, we clarify the critical property of the Anderson transition in the NH Peierls 
	model with the magnetic flux $2\pi \Phi=\pi/2$. 
	The localization length is calculated with 
	$L_z=10^7$ for $L=8,10,12,16,20$. The normalized localization length near 
	its scaling invariant point is shown in Fig.~\ref{pie}. The polynomial fitting results are 
	summarized in Table \ref{table}. For our choice of the data range, we need to take $n_1=4$ 
	to obtain a sufficiently large goodness of fit in the polynomial fitting. The 
	large $n_1$ represents a non-linear $\phi_1$-dependence of $F$ around the critical 
	point ($\phi_1=0$). The fittings with the system size $L=10\sim20$ and the expansion 
	order ($m_1$, $n_1$,  $m_2$,  $n_2$)=(2,4,0,1) or (3,4,0,1) give stable fitting results 
	with the critical exponent $\nu=1.02 \pm 0.01$. This value is quite close to 
	the critical exponent in the NH U(1) model, supporting Eq.~(\ref{CE2}) as the critical 
	exponent in the 3D NH class A systems. Note that the uniform magnetic 
	flux $\Phi$ introduces a geometric anisotropy in the Peierls model. 
	As a result, the critical normalized length $\Lambda_c$ in the NH Peierls model is 
	different from that in the NH U(1) model (Table \ref{table}), although the nearest-neighbor 
	hopping amplitudes in these two cubic-lattice models are all same in the three directions.

	\section{revisiting the level statistics analysis}
	\label{sec:RLSA}
	In this section, we compare the critical exponents 
	evaluated by the localization length with the previous evaluation by the level statistics 
	in the NH Anderson and U(1) models~\cite{Luo21}. 
	The level statistics analysis \cite{Wigner51,Dyson62,Dyson62TFW} introduces a finite energy window over which the 
	dimensionless quantity $\Gamma$ is averaged. When applying a linear fitting with respect to 
	$w\equiv (W-W_c)/W_c$, $\Gamma(W,L) =\Gamma_c + a w L^{1/\nu}$ 
	with $(m_1,n_1,m_2,n_2)=(1,1,0,0)$ in Eq.~(\ref{scaling}), the finite width of the energy 
	window does {\it not} cause any influence on the estimate of the critical exponent $\nu$; $a$ 
	and $W_c$ are averaged over the energy window, while $\Gamma_c$ and $\nu$ take the 
	same universal values for any states within the window. When applying a {\it non}-linear fitting of 
	$\Gamma$ with resepct to $w$, the analysis needs to assume that all the fitting parameters 
	in the universal function $F$ in Eq.~(\ref{scaling}) have no variations within the energy window. 
	As explained in Sec.~\ref{subsec:polynomiaFitting}, however, these fitting parameters include not 
	only the universal critical exponent but also non-universal critical quantities, such as critical disorder 
	strength $W_c$, and the scaling dimensions of the irrelevant scaling variables $y$. Generally, 
	two cases with different energies do not share the same non-universal critical quantities. 
	They are continuous functions of the energy. Thus, the energy window for the level statistics 
	must be narrow enough so that all the states inside the window share almost the same 
	values of these fitting parameters. Otherwise, the variations of the non-universal fitting 
	parameters cause additional systematic errors in the estimate of the universal critical exponent. 
	
	With the above consideration in mind, we first discuss the critical exponent of 
	the 3D class A models. The value obtained in this paper is consistent with the critical
	exponent of the NH U(1) model obtained by the level statistics analysis~\cite{Luo21}; 
	$\nu\approx 1.09$ by the level statistics for 10\% eigenvalues around $E=0$, and 
	$\nu\approx 1.01$ for 5\% eigenvalues around $E=0$. Namely, the critical exponent of the 3D class A
	models in this paper is closer to the value for the 5\% energy window than to the value for 
	the 10\% energy window. This tendency suggests that the level statistics analysis for the narrower energy 
	window are relatively free from the effects of finite variations of the non-universal critical quantities.
	
	On the one hand, the critical exponent of the 3D NH Anderson model (NH class AI$^\dagger$) obtained in this paper 
	is at variance with the critical exponent obtained by the level statistics analysis in Ref.~\onlinecite{Luo21}; 
	$\nu\approx 0.99$ by the level statistics for 10\% eigenvalues around $E=0$, and 
	$\nu\approx 0.95$ for 5\% eigenvalues around $E=0$. We regard that 
	the discrepancy comes from an interplay between the finite-energy window issue and a strong 
	asymmetry in a universal function for a level spacing ratio \cite{Oganesyan07,Huang20,Sa20} around the Anderson transition 
	point. The level spacing ratio $r$ exhibits limiting values \cite{Luo21,Sa20} as 
	a function of the disorder strength in the NH Anderson model; $\langle r \rangle_{\rm loc} = 2/3$ 
	in the localized phase and $\langle r\rangle_{\rm deloc} \approx 0.72$ in the delocalized phase. It turns 
	out that an intersection of curves of $\langle r \rangle$ for different system sizes,
	$\langle r\rangle_{\rm critical}= 0.716$, is 
	very close to one of the two limiting values,  $\langle r\rangle_{\rm deloc} \approx 0.72$.
	This indicates that the universal function for the level spacing ratio happens to be quite asymmetric 
	around the transition point in the NH Anderson model. When fitting such an asymmetric function by the 
	polynomial in $w$, the polynomial function must have a strong nonlinear $w$-dependence. Thus, 
	the finite-energy window issue together with this intrinsic strong non-linearity in the universal 
	function for the level spacing ratio causes large systematic errors in the NH Anderson model.
	In other words, a valid data range of polynomial fitting analysis with the smaller 
	$n_1$ and $m_1$ becomes very small in the side of the delocalized phase and the choice 
	of the data points in the previous study is not narrow enough in the NH Anderson model. 
	
	In summary, the argument so far concludes the critical exponent $\nu=1.19\pm 0.01$ 
	in the Anderson transition in the NH symmetry class AI$^{\dagger}$, and 
	$\nu=1.00 \pm 0.04$ in the Anderson transition in the NH symmetry class A. 
	In spite of the correction of the critical exponent 
	in the NH class AI$^{\dagger}$, our previous conclusion in the 
	level statistics analysis remains unchanged. Namely, 3D NH Anderson transition 
	in classes AI$^\dagger$ and in class A show critical behaviors different from 
	their Hermitian counterparts, hence belong to new universality classes.

	\section{Summary}
	\begin{table*}[tbh]
		\centering
		\setlength{\tabcolsep}{1.1mm}
		
		\begin{tabular}{|c|c|c|c|c|c|c|c|}
			\hline
			Model&class &$H$&$M$& Lyapunov & $S$ & unitarity & $g$  \\
			\hline\hline
			H AM & AI & $H=H^\mathrm{T}=H^*$&$\sigma_y M^\mathrm{T} \sigma_y=M^{-1}$&$\gamma_i=\gamma_i'$&$S=S^\mathrm{T}$&$S^\dagger S=I$&$g_L=g_R$\\
			\hline
			H U & A & $H=H^\dagger$ &$\sigma_y M^\dagger \sigma_y=M^{-1}$&$\gamma_i=\gamma_i'$&--& $S^\dagger S=I$&
			$g_L(\pm \{\phi\})=g_R(\pm \{\phi\})$\\
			\hline
			NH AM& NH AI$^\dagger$& $H=H^\mathrm{T}$&$\sigma_y M^\mathrm{T} \sigma_y=M^{-1}$&$\gamma_i=\gamma_i'$&$S=S^\mathrm{T}$&$S^\dagger(H) S(H^\dagger)=I$ & $g_L=g_R$\\
			\hline
			NH U& NH A & --&--&$\langle \gamma_i\rangle=\langle\gamma_i'\rangle$&--&$S^\dagger(H) S(H^\dagger)=I$ & $g_L(\{\phi\})=g_R(-\{\phi\})$\\
			\hline
			p-H &   & $\eta H\eta^{-1}=H^\dagger$&
			$\eta\sigma_y M^\mathrm{\dagger}(E)\sigma_y\eta^{-1}=M^{-1}(E^*)$&
			$\gamma_i(E)=\gamma_i'(E^*)$& &$S^\dagger(E)\eta S(E^*)=\eta$ & $g_L=g_R$\\
			\hline
		\end{tabular}
		\caption{Summary of symmetry relations of Lyapunov exponents (`Lyapunov'), the $S$ matrix 
			($S$) and two-terminal conductances ($g$) in Sec.~\ref{sec:method}. The columns of $H$ and $M$ specify the 
			symmetries of Hamiltonian and transfer matrix in the Hermitian Anderson model (H AM), 
			non-Hermitian Anderson model (NH AM), Hermitian U(1)/Peierls model (H U) and non-Hermitian 
			U(1)/Peierls model (NH U). `$\gamma_{i}=\gamma^{\prime}_{i}$' in the column of `Lyapunov' 
			stands for a $\pm$ symmetry of the Lyapunov exponents: the Lyapunov exponents 
			are ordered such that $\gamma_n>\cdots\gamma_2> \gamma_1>0>-\gamma_1'>-\gamma_2'\cdots>-\gamma_n'$, $2n=2L_x L_y$ being the dimensions of the transfer matrix. In the non-Hermitian case, a Hermitian conjugate of 
			a $S$ matrix for $H$ is an inverse of a $S$ matrix of $H^{\dagger}$; $S^{\dagger}(H) = S^{-1}(H^{\dagger})$.  
			For the NH U(1)/Peierls models, the leftward conductance with a position dependent magnetic flux $\{\phi\}$ 
			is identical to the rightward conductance with $-\{\phi\}$; $g_L(\{\phi\})=g_R(-\{\phi\})$. For the Hermitian 
			U(1)/Peierls models, the unitarity of the scattering matrix gives an equivalence between the leftward conductance 
			and rightward conductance; $g_L(\{\phi\})=g_R(\{\phi\})=g_L(-\{\phi\})=g_R(-\{\phi\})$.
			`p-H' indicates non-Hermitian systems with the pseudo-Hermiticity in Sec.~\ref{appendix_Pseudo}. $\eta$ is a Hermitian and unitary matrix.}
		\label{tab:symm}
	\end{table*}

	In this paper, we presented transfer matrix analyses of the Anderson transition in the 
	three NH systems with on-site complex-valued random potentials that belong to the 
	class-AI$^{\dagger}$ (NH Anderson model) and class-A (NH U(1) and NH Peierls models),
	respectively. We first provided an argument with solid numerical evidence that supports the validity of 
	the transfer matrix analyses of the localization length and two-terminal conductance in NH systems. 
	We then clarified the presence or absence of the reciprocal symmetries of the Lyapunov exponent 
	and the conductance in the three NH models.  The relations are summarized in Table \ref{tab:symm}.
	We note that other relations in non-reciprocal non-Hermitian 
	systems are found in \cite{Schomerus21}.

	On the basis of the above knowledge, we evaluated the critical 
	exponents at different single-particle energies and different system parameters of the NH Anderson 
	model as well as of the NH U(1) and NH Peierls models. The results conclude that the critical exponent of 
	the NH class-AI$^{\dagger}$ is $\nu=1.19 \pm 0.01$ and the critical exponent of the NH class-A 
	is $\nu=1.00 \pm 0.04$, indicating a strong linkage between the universality class of the Anderson 
	transition and symmetry classification in the NH systems. From the critical properties 
	at different system parameters of the NH Anderson model, we draw a 
	phase diagram in a two-dimensional plane subtended by the disorder strength of 
	the real part of the on-site random potential ($W_r$) and that of the imaginary part ($W_i$). 
	We showed that at the zero single-particle energy ($E=0$), the phase diagram 
	as well as the critical properties become completely symmetric with respect to an exchange 
	between  $W_r$ and $W_i$. We further proved that the symmetric structure at the zero single-particle 
	energy is a generic feature in any NH bipartite-lattice models with the on-site NH random potentials. 
	We also demonstrated numerically that a distribution of the two-terminal 
	conductance is not Gaussian and the distribution contains small numbers of huge conductance, 
	which come from rare events of strong amplifications of the transmission by the NH disorders. 
	By analyzing the geometric mean of the conductance, we show that the critical properties of 
	the conductance in the NH Anderson and NH U(1) models also give critical exponents, which are 
	consistent with those critical exponents obtained by the localization length.
	We note that Wegner's relation\cite{Wegner76}, which relates the conductivity critical exponent $s$ and 
	the localization critical exponent $\nu$ as $s=(d-2)\nu=\nu$, remains to be checked in NH systems.
	
	In conclusion, an experimental verification of the new universality classes 
	in the NH systems may be possible in disordered optical systems~\cite{John87,Wiersma97,Schwartz07,Skipetrov15,Skipetrov18}
	and acoustic systems~\cite{Kirkpatrick85,WEAVER90,Hu08}, which usually 
	fall into the 3D NH class AI$^\dagger$. The energy loss and gain in the optical 
	systems make it difficult to verify the Anderson transition in the Hermitian 
	systems. On the other hand, disordered media with gain and loss are 
	ideal platforms for an experimental verification of the Anderson transition 
	in NH systems~\cite{Tzortzakakis20,Tzortzakakis21}. 
	An experimental realization of the NH class A model in the disordered optical systems 
	seems to be non-trivial, and we leave it for an important future problem. 
	
	%

	\bigskip
	\bigskip
	
	\noindent
	\begin{acknowledgments}
		The authors thank the fruit discussions and comments with Prof. Nariyuki Minami, 
		Prof. Keith Slevin and Dr. Kohei Kawabata. X. L. was supported by National Natural
		Science Foundation of China of Grant No.51701190. T. O. was supported by 
		JSPS KAKENHI Grants No. 16H06345 and 19H00658. R. S. was supported by 
		the National Basic Research Programs of China (No. 2019YFA0308401) and by 
		National Natural Science Foundation of China (No.11674011 and No. 12074008). 
	\end{acknowledgments}
	
	\appendix*

		\section{Two-terminal conductance and its finite-size scaling analysis}
	
	\begin{figure}[t]
		\centering
		\includegraphics[width=1.1\linewidth]{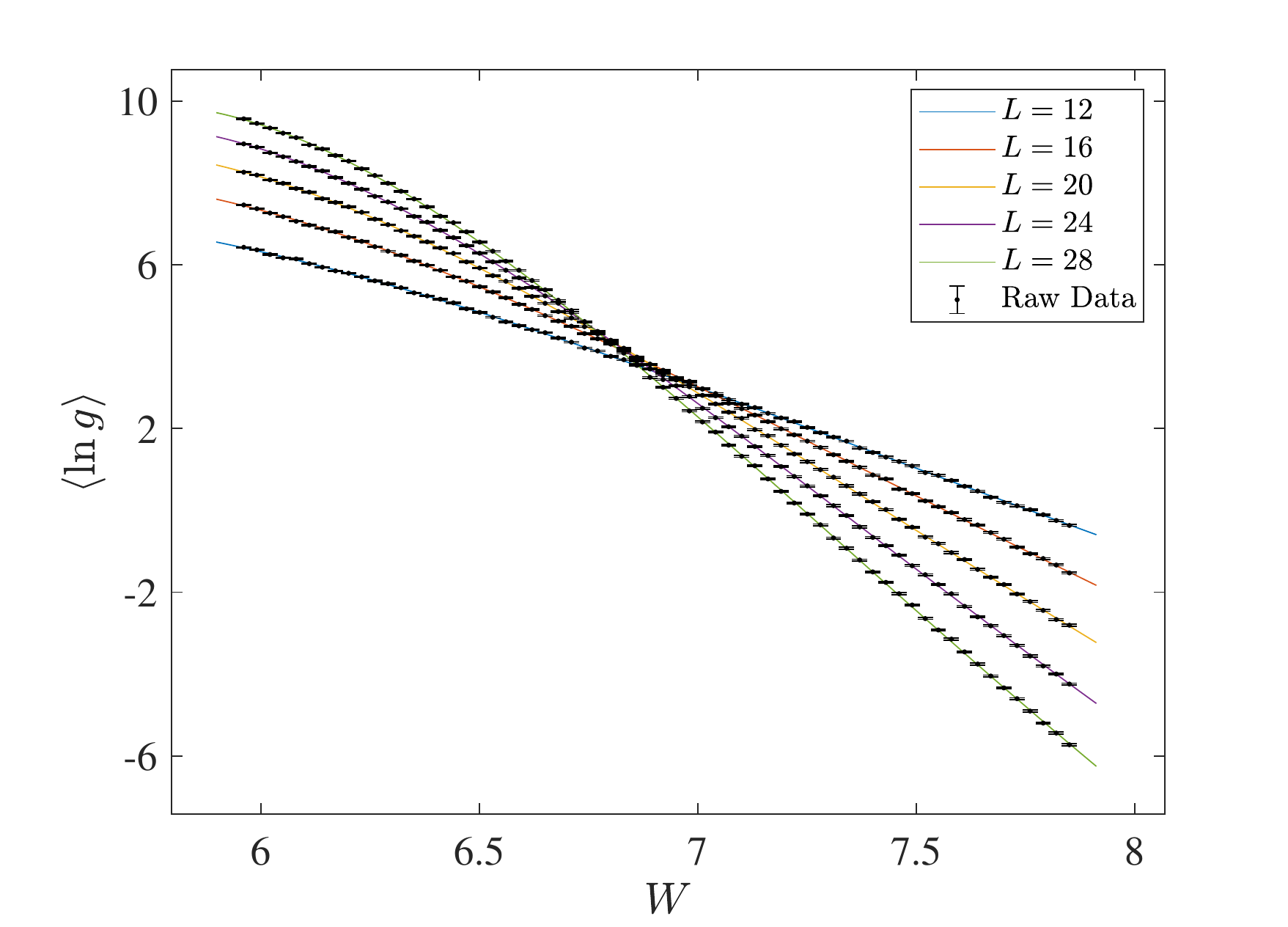}
		\caption{Polynomial fitting for $\langle\ln g\rangle$ at $E=0$ with $W_r=W_i=W$ for the NH 
			Anderson model with periodic boundary condition in the $x$ and $y$ directions. 
			The black points with error bar are the raw data of $\langle\ln g\rangle$ and lines with different colors  
			are the polynomial fitting results with the expansion order $(m_1, n_1,  m_2,  n_2)=(2,3,0,1)$. 
			$\langle\ln g\rangle$ are averaged over $10^4$ samples. Note also that the reciprocal relation 
			of the conductance is verified numerically in each sample.}
		\label{lng_AM}
	\end{figure}

	\begin{figure*}[t]
		\centering
		\subfigure[ PBC, (3,3,0,1) ]{
			\begin{minipage}[t]{0.5\linewidth}
				\centering
				\includegraphics[width=1\linewidth]{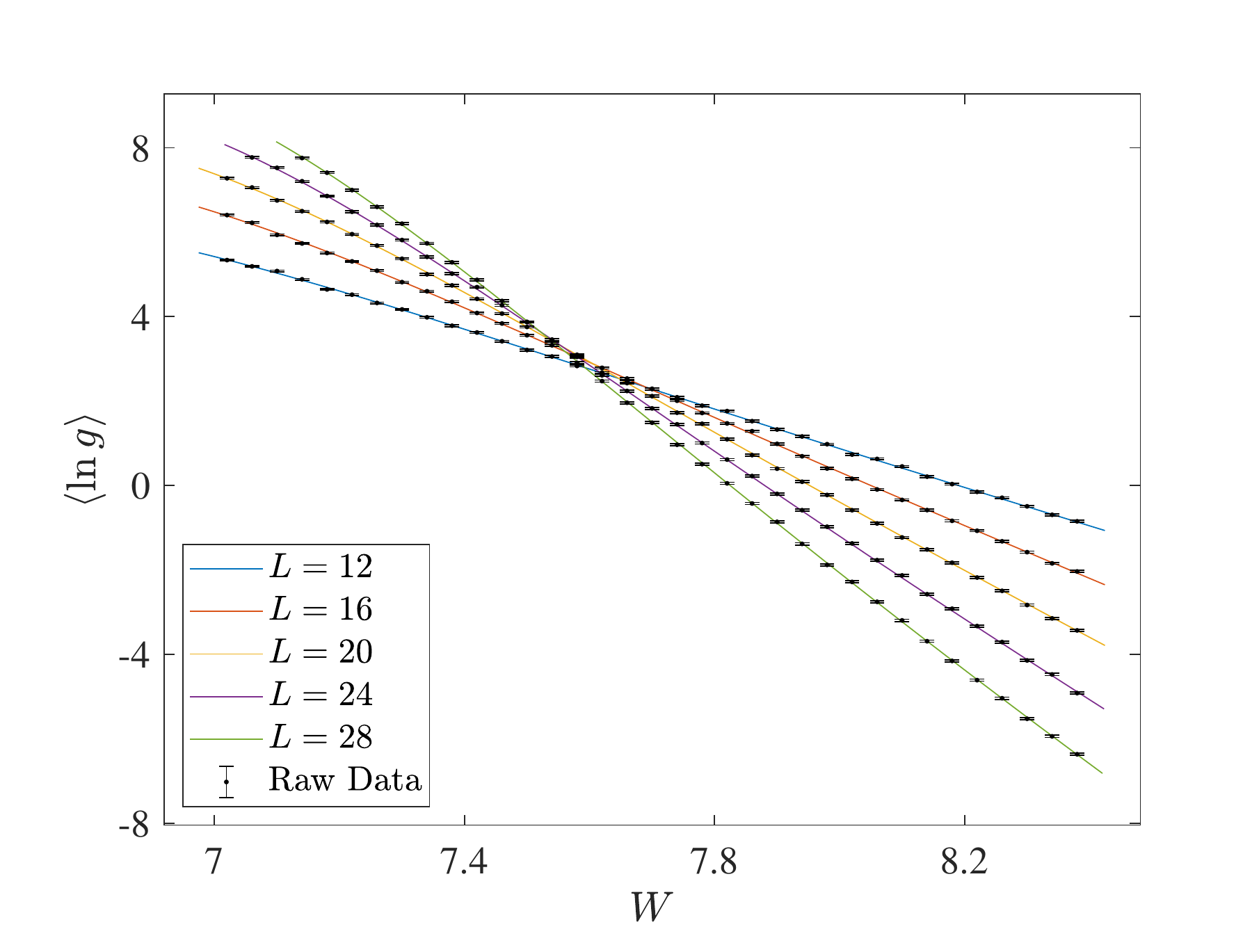}
			\end{minipage}%
		}%
		\subfigure[ OBC, (3,3,0,1) ]{
			\begin{minipage}[t]{0.5\linewidth}
				\centering
				\includegraphics[width=1\linewidth]{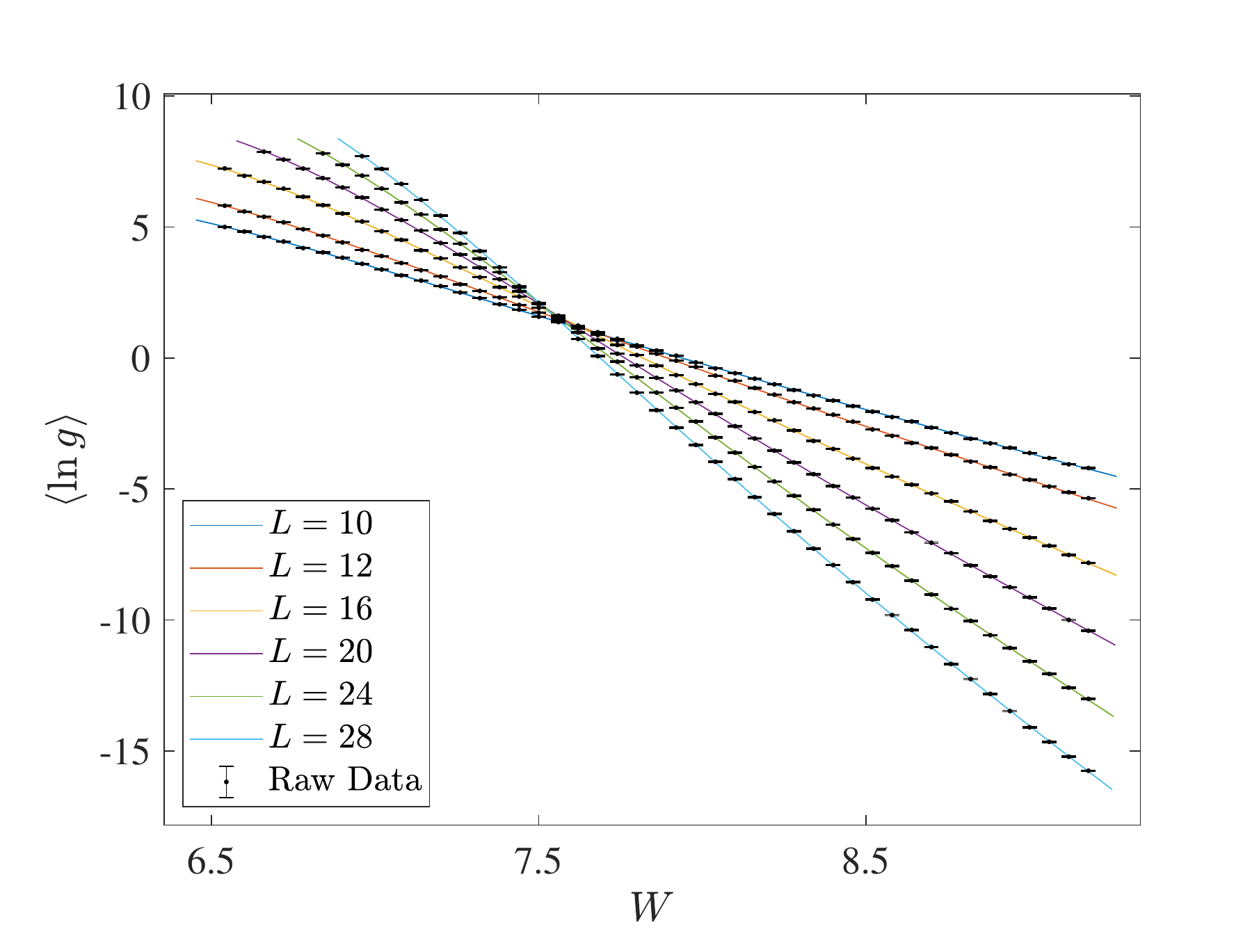}
			\end{minipage}%
		}%
		\caption{Polynomial fitting for $\langle\ln g\rangle$ at $E=0$ with $W_r=W_i=W$ for the NH U(1) model with the 
			(a) periodic boundary condition (PBC) and (b) open boundary condition (OBC). The black points with error bar are the raw data 
			of $\langle\ln g\rangle$ and lines with different colors are the polynomial fitting results with  expansion 
			order $(m_1, n_1,  m_2,  n_2)$.  
			The expansion order ($m_1$, $n_1$,  $m_2$,  $n_2$) are shown in the brackets for each figure.
			$\langle\ln g\rangle$ are averaged over $10^4$ samples.}
		\label{lng_U1}
	\end{figure*}

	\begin{table*}[t]
		\setlength{\tabcolsep}{3.5mm}
		\caption{Polynomial fitting results for the logarithm of the conductance $\langle\ln g\rangle$ 
			around the Anderson transition points in the NH Anderson and U(1) models at $W_r=W_i=W$ 
			and $E=0$. The goodness of fit (GOF), critical disorder $W_{c}$, critical exponent $\nu$, the 
			scaling dimension of the least irrelevant scaling variable $-y$ are shown for various system sizes 
			and for different expansion orders of the polynomial: $(m_1,n_1,m_2,n_2)$. 
			The square bracket stands for the 95\% confidence interval.}
		\begin{tabular}{ccccccccc}
			&&&&&&&&\\
			\multicolumn{9}{l}{NH Anderson model, with the PBC}\\
			\hline
			$L$&$m_1$& $n_1$ & $m_2$ & $n_2$ & GOF & $W_c$ & $\nu$ & $y$  \\
			\hline
			12-28&2&3&0&1&0.20&6.415[6.387, 6.523]&1.276[0.998, 1.454]&0.06[0.03, 0.32]\\
			12-28&3&3&0&1&0.23&6.415[6.385, 6.513]&1.288[1.006, 1.524]&0.06[0.03, 0.30]\\
			16-28&2&3&0&1&0.20&6.415[6.387, 6.520]&1.276[1.002, 1.457]&0.06[0.03, 0.31]\\
			16-28&3&3&0&1&0.23&6.415[6.385, 6.522]&1.288[1.006, 1.520]&0.06[0.03, 0.31]\\
			\hline
			&&&&&&&&\\
			\multicolumn{9}{l}{NH U(1) model, with the PBC}\\
			\hline
			$L$&$m_1$& $n_1$ & $m_2$ & $n_2$ & GOF & $W_c$ & $\nu$ & $y$  \\
			\hline
			12-28&	3&3&0&1&0.10&7.338[7.323, 7.435]&1.142[0.861, 1.349]&0.044[0.024, 0.639]\\
			16-28&	3&3&0&1&0.10&7.329[7.295, 7.362]&1.150[0.859, 1.479]&0.046[0.039, 0.091]\\
			\hline
			&&&&&&&&\\
			\multicolumn{9}{l}{NH U(1) model, with the OBC}\\
			\hline
			$L$&$m_1$& $n_1$ & $m_2$ & $n_2$ & GOF & $W_c$ & $\nu$ & $y$  \\
			\hline
			10-28&3&3&0&1&0.24&7.391[7.379, 7.444]&1.069[0.961, 1.130]&0.12[0.07, 0.55]\\
			12-28&3&3&0&1&0.12&7.397[7.381, 7.454]&1.075[0.950, 1.162]&0.15[0.08, 0.70]\\
			\hline
		\end{tabular} \label{table_g}
	\end{table*}
	
	In this appendix, we describe behaviours of the two-terminal conductance as a 
	function of disorder strength and the FSS analyses of the conductance 
	for the NH Anderson and U(1) models.
	
	Fig.~\ref{lng_AM} shows the logarithm of the conductance $g$ 
	as a function of $W$ in the NH Anderson model at $W_r=W_i=W$ and $E=0$.
	Thereby, the conductance $g$ is calculated in terms of 
	Eqs.~(\ref{two-terminal-conductance},\ref{whatisT},\ref{T-def}) with 
	the cubic geometry $L_x=L_y=L_z=L$ and periodic boundary condition (PBC) 
	along the $x$ and $y$ direction.  The logarithm of $g$ is averaged 
	over $10^4$ samples of different disorder realizations. 
	$\langle\ln g\rangle$ for different system size $L$ show an intersection near the critical 
	disorder strength determined by the Lyapunov exponent analysis. Nonetheless, 
	the intersection point of $\langle \ln g \rangle$ for different system sizes 
	is not as obvious as in the Lyapunov exponent, even for larger system size 
	($L=16,20,24,28$); $\langle\ln g\rangle$ among larger 
	system sizes tends to intersect at a smaller disorder strength. 
	
	The critical exponent is obtained from the polynomial fitting analysis 
	(Table \ref{table_g}). The critical exponent evaluated from $\langle \ln g \rangle$ is 
	slightly larger than the critical exponent determined by the Lyapunov 
	exponent, while they are consistent with each other within the 95\% 
	confidence intervals. Note that the scaling dimension of the least irrelevant 
	scaling variable is quite small in the polynomial fitting analysis of 
	$\langle \ln g\rangle$, which results in a large error bar for $\nu$. 
	The poor estimation of the critical property comes from the poor intersection 
	of $\langle\ln g\rangle$. Two reasons could be responsible for the poor 
	intersection; (i) a non-optimal choice of the contact between the Hermitian leads 
	and the NH system and (ii) intrinsically large values of the conductance due to 
	the non-Hermiticity.
	
	Fig.~\ref{lng_U1} shows plots of the logarithm of the 
	conductance as a function of $W$ in the NH U(1) model at 
	$W_r=W_i=W$ and $E=0$ with the same conditions as above, 
	except for the boundary condition along the $x$ and $y$ direction. 
	The plot with the PBC is in Fig. \ref{lng_U1}(a) and the plot with 
	the open boundary condition (OBC) is in Fig.~\ref{lng_U1}(b). The polynomial fitting result is summarized 
	in Table \ref{table_g}. As in the NH Anderson model, the critical disorder 
	strength from the fitting is slightly larger than the critical 
	disorder strength determined by the Lyapunov exponent and the intersection point 
	shifts to a smaller value for larger system sizes. Nonetheless, the critical exponent 
	is consistent with the critical exponent determined by the Lyapunov exponent 
	within the 95\% confidence interval. Note that the smaller value of $y$ 
	from the fitting analysis indicates a strong finite-size correction, which leads to a 
	large error bar of the critical exponent and the poor intersection.
	Note also that the intersection of the conductance with the OBC is better than 
	that with the PBC. This phenomenon doesn't occur for the NH Anderson model.
	
	\bibliography{paper}
\end{document}